\documentclass[article,shortnames]{jss}

\usepackage{orcidlink,thumbpdf,lmodern}

\usepackage{framed}

\usepackage{tikz}
\usetikzlibrary{arrows}
\usetikzlibrary{fit,positioning}
\usepackage{amsmath,bm,amssymb}
\usepackage{graphicx,psfrag,epsf}
\usepackage[utf8]{inputenc}
\usepackage{enumitem}
\newcommand{\class}[1]{`\code{#1}'}
\newcommand{\fct}[1]{\code{#1()}}
\newcommand{\rmn}[1]{{\mathrm{#1}}}
\newcommand{\nc}{\newcommand}
\nc{\tb}[1]{\textcolor[rgb]{0.00,0.00,1.00}{#1}}
\nc{\trd}[1]{\textcolor[rgb]{1.00,0.00,0.00}{#1}}
\nc{\tp}[1]{\textcolor[rgb]{1.00,0.00,1.00}{#1}}
\newcommand{\indep}{\perp \!\!\! \perp}
\nc{\trans}{^{\top}}

\author{ Li Su~\orcidlink{0000-0003-0919-3462}, Roonak Rezvani~\orcidlink{0000-0001-5580-5058}, Shaun R. Seaman~\orcidlink{0000-0003-3726-5937}, Colin Starr\\ University of Cambridge
     \AND Isaac Gravestock~\orcidlink{0000-0003-0283-2065}\\F.Hoffmann - La Roche Ltd}
\Plainauthor{Li Su, Roonak Rezvani, Shaun R. Seaman, Colin Starr, Isaac Gravestock}

\title{\pkg{TrialEmulation}: An \proglang{R} Package to Emulate  Target Trials for Causal Analysis of Observational Time-to-event Data}
\Plaintitle{TrialEmulation: An R Package to Emulate  Target Trials for Causal Analysis of  Observational Time-to-event Data}
\Shorttitle{\pkg{TrialEmulation}: Emulating Target Trials in \proglang{R}}

\Abstract{
Randomised controlled trials (RCTs) are regarded as the gold standard for estimating causal treatment effects  on health outcomes.
However, RCTs are not always feasible, because of time, budget or ethical constraints.
Observational data such as those from electronic health records (EHRs) offer an alternative way to estimate the causal effects of treatments.  Recently,
the `target trial emulation' framework was proposed by \cite{Hernan2016} to provide a formal structure for estimating causal treatment effects from observational data.
To promote more widespread implementation of target trial emulation in practice, we develop the \proglang{R} package \pkg{TrialEmulation} to emulate a sequence of target trials using observational time-to-event data,  where individuals who start to receive treatment and those who have not been on the treatment at the baseline of the emulated trials are compared in terms of their risks of an outcome event.  Specifically, \pkg{TrialEmulation} provides (1) data preparation for emulating a sequence of target trials, (2) calculation of the inverse probability of treatment and censoring weights to handle treatment switching and dependent censoring, (3)  fitting of marginal structural models for the time-to-event outcome given baseline covariates,  (4)  estimation and inference of marginal \emph{intention to treat} and \emph{per-protocol} effects of the treatment in terms of marginal risk differences between treated and untreated for a user-specified target trial population.    In particular, \pkg{TrialEmulation}  can accommodate large data sets (e.g., from EHRs) within memory constraints of \proglang{R} by processing data in chunks and applying case-control sampling.  We demonstrate the functionality of \pkg{TrialEmulation} using a  simulated data set that mimics typical observational time-to-event data in practice.}

\Keywords{big data, causal inference, censoring, inverse probability weighting,  survival data}

\Address{
  Li Su, Roonak Rezvani, Shaun R. Seaman, Colin Starr\\
  MRC Biostatistics Unit\\
  School of Clinical Medicine\\
  University of Cambridge\\
  Forvie site, Robinson Way\\
  Cambridge CB2 0SR, UK\\
  E-mail:
  \\\email{li.su@mrc-bsu.cam.ac.uk;
  \\~~~~~~~~~~roonak.r74@gmail.com;\\ shaun.seaman@mrc-bsu.cam.ac.uk\\
  colin.starr@mrc-bsu.cam.ac.uk}\\

Isaac Gravestock\\
Real World Data, PD Data Sciences\\
F.Hoffmann - La Roche Ltd\\
E-mail: \email{isaac.gravestock@roche.com}

 }

\begin{document}



\section{Introduction}\label{intro}

Randomised controlled trials (RCTs) are regarded as the gold standard for estimating causal treatment effects on health outcomes. By randomising individuals to receive treatment or non-treatment, these trials ensure that the treated group of individuals and the untreated group of individuals are comparable at the time of treatment assignment.  However, RCTs are not always feasible, because of time, budget or ethical constraints, or have not been used to study particular populations, e.g., RCTs often exclude individuals with co-morbidities.

Observational data such as those from electronic health record (EHR) databases offer an alternative way to estimate causal treatment effects.  However, valid estimation of causal treatment effects from such databases is not straightforward.  For example, because treatment assignment in EHR databases is usually not randomised, a difference between the means of the observed outcomes in treated and untreated groups of individuals may be at least partly explained by baseline differences between the groups, rather than by the treatment effect  (i.e., the treatment effect is confounded).  Recently, the  `target trial emulation' framework was proposed by \cite{Hernan2016} to provide a formal structure for estimating causal treatment effects using observational data.  This involves specifying the protocol of a `target trial' that one would like to carry out, identifying the individuals in an observational database who satisfy the eligibility criteria of this target trial, and then comparing the outcomes between treated and untreated eligible individuals during the trial follow-up period.  The lack of randomisation in treatment assignment can be at least partly accounted for by using techniques such as matching, weighting or regression adjustment.  Other sources of bias that may need to be addressed include treatment switching and dependent censoring after the trial baseline.

To promote more widespread implementation of target trial emulation in practice,  we develop the \pkg{TrialEmulation}  package  \citep{Rezvani2022}, which,  to the best of our knowledge, is the first \proglang{R} package dedicated to target trial emulation. Specifically, \pkg{TrialEmulation}
 provides a causal analysis pipeline of emulating a sequence of target trials, in each of which the individuals who started to receive (`initiated') treatment at the trial baseline are compared in terms of the risk of an outcome event with the individuals who hadn't initiated the treatment.
 This `sequential trials' approach was first described in \cite{Hernan2008} and \cite{Gran2010}  as a simple way to efficiently analyse observational longitudinal data for causal effect estimation.  This approach recognises that individuals could meet the trial eligibility criteria multiple times during their follow-up in an observational database and thus emulates a sequence of target trials, each starting at a different time.  The analyses of these trials are then combined to produce an overall estimate of the treatment effect. This approach is attractive when the number of individuals who met the trial eligibility criteria and initiated treatment at any time is small and/or the number of outcome events that occurred in a single trial is small \citep{Danaei2013}. \cite{Keogh2021} provided a recent review of the sequential trials approach and its connections with the alternative marginal structural model approach \citep{Robins2000}.

There are several ways to prepare data and perform causal analyses of observational time-to-event data in a sequence of emulated trials  (see the review by \citealp{Keogh2021}).  \pkg{TrialEmulation} largely follows the  methodology described in \cite{Danaei2013}, which was  implemented in a \proglang{SAS} \citep{SAS-STAT} macro \pkg{initiators} \citep{Logan2011}.
 Specifically,
\pkg{TrialEmulation} takes an observational data set provided by the user and expands it to create the data set corresponding to a sequence of target trials such that both \emph{intention-to-treat} and \emph{per-protocol} analyses can be performed. We will provide details of this data expansion procedure in Section~\ref{section2}.
\pkg{TrialEmulation} calculates the inverse probability of censoring weights to account for censoring of follow-up, e.g., due to individuals' withdrawal from the study.
In addition, for per-protocol analyses, where follow-up is artificially censored when treatment switching occurs, \pkg{TrialEmulation} calculates inverse probability weights that correct for the selection bias that would otherwise result from this artificial censoring.
\pkg{TrialEmulation} then fits a marginal structural model for the time-to-event outcome of interest by applying a  pooled logistic regression with these inverse probability weights to the expanded data set.
The non-random treatment assignment at baseline is accounted for by including variables measured at the trial baselines as covariates in the pooled logistic regression model.
This pooled logistic regression model is an approximation to a
proportional hazards model, and the baseline hazard function can be flexibly modelled by the use of splines \citep{Hernan2000,Murray2021}.  The standard errors of the parameters in the pooled logistic regression model are estimated by the robust sandwich estimator, in order to account for the weights and for correlations induced by data from the same individual being used in multiple trials \citep{Hernan2000,Danaei2013}.

An advantage of \pkg{TrialEmulation} compared to the \proglang{SAS} macro \pkg{initiators} is that the former provides estimation and inference of marginal treatment effects in terms of risk differences between treated and untreated for a user-specified target trial population. This is attractive because marginal risk difference is a commonly used causal estimand in  RCTs with time-to-event outcomes \citep{Young2020}.  Specifically, after the user has provided baseline covariate values for each individual in a target trial population, \pkg{TrialEmulation} predicts (using the fitted pooled logistic regression model) the cumulative incidence 
for each individual twice: once when treatment is received and once when it is not.
The marginal risk difference is then estimated by averaging the differences between these pairs of predicted cumulative incidences \citep{Murray2021}.  A simulation-based confidence interval \citep{Mandel2013} for the marginal risk difference is constructed by sampling from the asymptotic distribution of the parameters in the pooled logistic regression model and making repeated predictions of cumulative incidences.

Another improvement
of \pkg{TrialEmulation} over the \proglang{SAS} macro \pkg{initiators} is that it provides functionality to accommodate large data sets.
If the original data set provided by the user contains a large number of individuals, then the expanded data set, which contains multiple copies of each individual, will be very large.  Processing such a large amount of data in one go is prohibitive, due to the memory constraints in \proglang{R}. \pkg{TrialEmulation} provides the option of processing the data in chunks and then combining before the analysis is carried out.
Fitting the pooled logistic regression model to the entire expanded data set and calculating the sandwich variance estimate will also involve a very heavy computational load.
\pkg{TrialEmulation}  offers the option of case-control sampling to reduce the volume of the expanded data set (this is described in Section~\ref{sampling}).

In this paper, we provide a comprehensive illustration of the \pkg{TrialEmulation} package to assist researchers in performing intention-to-treat and per-protocol analyses for a sequence of emulated trials. In Section~\ref{section2}, we describe the target trial emulation method that is implemented in \pkg{TrialEmulation}. Section~\ref{pkgdetails} outlines some important technical details and the main functions in \pkg{TrialEmulation}. Section~\ref{sec:illustrations} illustrates the use of these functions with a data set simulated based on  a typical observational time-to-event data setting with treatment switching and dependent censoring \citep{Young2014}. We conclude in Section~\ref{sec:summary} with a discussion.



\section{Overview of target trial emulation}\label{section2}

In this section, we describe the methodology of emulating a sequence of target trials using observational time-to-event data that is implemented in \pkg{TrialEmulation}. As the sequential trials setting introduces complexities in notation and implementation of trial emulation, we start with the single trial setting to facilitate understanding.

\subsection{Emulating a single target trial}\label{singletrial}

\subsubsection{Setting and notation}\label{observed}

To emulate a target trial using observational data, we first need to specify the protocol of the target trial. Table~\ref{protocol} describes the protocol of a hypothetical trial to estimate the effect of an antiglycaemic drug on the 5-year risk of cardiovascular events among type-2 diabetes patients.

\begin{table}[ht!]
\centering
\caption{\label{protocol} A summary of the protocol of a target trial to estimate the effect of an antiglycaemic drug on the 5-year risk of cardiovascular events among type-2 diabetes patients}
\begin{tabular}{lp{10cm}}
\hline
\hline
\textbf{Protocol component }  & \textbf{Description} \\ \hline
Eligibility criteria      & Type 2 diabetes patients, $\ge 45$ years old
							with sub-optimal glycemic control
  between the years 2000-2020, with no history of using the antiglycaemic drug in the past two years and no history of cardiovascular events. \\
  \\
 Treatment strategies    & 1) Non-treatment:  refrain from taking the antiglycaemic drug during the follow-up. 2) Treatment: initiate the antiglycaemic drug at baseline and continue to receive it during the follow-up unless it is discontinued due to side effects. \\ 
 \\
   Assignment procedures & Patients will be randomly assigned to either strategy at baseline and will be aware of the strategy to which they have been assigned.\\
   \\

   Follow-up period      &  Starts at randomisation and ends at the occurrence of a cardiovascular event, death, loss to follow-up or 5 years after baseline, whichever occurs first. \\
   \\
Outcome       & Occurrence of a cardiovascular event or death. \\
\\
Causal estimands      & Intention-to-treat effect, per-protocol effect. \\
\\
 Analysis plan      & Intention-to-treat analysis: estimation is via the comparison of the outcome risks among patients assigned to each treatment strategy, where pre-baseline covariates are adjusted for using regressions. Per-protocol analysis: estimation also accounts for post-baseline covariates associated with adherence to the treatment strategies using inverse probability of treatment weighting. Both analyses will account for  post-baseline covariates associated with loss to follow-up using inverse probability of censoring weighting.  Marginal treatment effects are then obtained by averaging over the
 pre-baseline covariate distributions of the trial population and comparing the averaged outcome risks under treatment strategies among the trial participants.  \\ \hline
\end{tabular}

\end{table}

Suppose that we have identified $n$ individuals from an observational database who met the eligibility criteria in the target trial protocol. Each of the $i=1, \ldots, n$ individuals were followed from the trial baseline at time $t_0$. Individuals were assumed independent and identically distributed in
the emulated trial and thus we suppress the individual-specific subscript $i$.
Longitudinal
measurements from individuals were taken at regular visits $k = 0, \ldots, K$ at time $t_0, \ldots, t_K$, where $t_K$ was the maximum follow-up time of the emulated trial (e.g., 5 years since baseline).  Let $\mathbf{V}$ be a vector of time-invariant covariates such as gender and race/ethnicity.  Let $L_k$ denote a
vector of time-varying covariates measured at time $t_k$,  so  $L_0$ denotes baseline values
of the time-varying covariates. Note that, when time-varying covariates were not measured for all individuals at all visits $k = 0, \ldots, K$, $L_{k}$ may contain the most recently measured values of the individual's time-varying covariates and the time since the last measurement \citep{Hernan2009}.
Let $A_k$ denote the observed treatment status of an individual at time $t_k$,
where $A_k = 1$ ($A_k=0$) means the individual was receiving (was not receiving) treatment at time $t_k$.
Let $Y_{k}$ be the binary variable indicating whether the event of interest occurred during the interval $[0, t_{k+1})$.
In practice, loss to follow-up is likely to occur for
some individuals, and this may depend on received treatments, time-invariant and time-varying covariates.  Let  $C_k = 1$ indicate that the individual is lost to follow-up in the interval
$[t_k, t_{k+1})$ (and so $Y_{k+1}, \ldots,  Y_K$ are missing), and let $C_k = 0$ otherwise. We assume
the time ordering $(L_k,A_k, Y_k, C_k)$.
 We use an overbar to denote the history of a random variable; for example, $\overline{A}_k=(A_0, \ldots, A_k)$ is the treatment history by $t_k$.

To describe the causal estimands in the next section, we 
 define the potential outcome variables $Y_{k}^{a}$  ($a=0,1$)  as follows. For each individual, $Y_{k}^{a}$ is the indicator of whether the event of interest would have occurred during the time interval $[0, t_{k+1})$ if the individual, possibly contrary to fact, had been assigned to the treatment $a$ at baseline.
In addition, for each individual, $Y_{k}^{\overline{a}_k}$ is the indicator of whether the event of interest would have occurred during the time interval $[0, t_{k+1})$,  if the individual, possibly contrary to fact, had taken treatment $\overline{a}_k$ up to time $t_k$.  For example, two values of $\overline{a}_k$, viz.\ $\overline{a}_k= (1, \ldots, 1)$ or $\overline{a}_k= (0, \ldots, 0)$ correspond to starting treatment/non-treatment at baseline and adhering to this treatment/non-treatment until at least time $t_k$.

\subsubsection{Causal estimands and assumptions}\label{estimands}

\pkg{TrialEmulation} provides two options of causal estimands, that is, the causal effects that the user wishes to estimate.

\begin{enumerate}
\item[1)] \emph{Marginal intention-to-treat effect}. This is the marginal effect of  \emph{treatment randomisation} on the cumulative incidence of the outcome event over the follow-up time since the trial baseline.  In \pkg{TrialEmulation}, this effect is defined as the difference between the cumulative incidences  when all individuals in the target trial population are assigned at baseline to treatment and non-treatment, respectively: 
\begin{equation}\label{ITTestimand}
     \mbox{Pr}(Y_{k}^{a=1}=1)-\mbox{Pr}(Y_{k}^{a=0}=1).
\end{equation}

\item[2)] \emph{Marginal per-protocol effect}.  This is the marginal effect of  \emph{sustained treatments} on the cumulative incidence over the follow-up time since the trial baseline. In \pkg{TrialEmulation}, this effect is defined as the difference between the cumulative incidences  when all individuals in the target trial population  remain on the treatment assigned and the cumulative incidences when all individuals remain untreated after assignment to the non-treatment at baseline, 
\begin{equation}\label{PPestimand}
      \mbox{Pr}(Y_{k}^{\overline{a}_k=(1, \ldots, 1)}=1)-\mbox{Pr}(Y_{k}^{\overline{a}_k=(0, \ldots, 0)}=1).
\end{equation}
\end{enumerate}

Identification of the intention-to-treat effect in~\eqref{ITTestimand} requires four key assumptions: no interference, consistency, positivity and no unmeasured confounding of treatment assignment at the trial baseline.
In addition to the four key assumptions, identification of the per-protocol effect in~\eqref{PPestimand} requires positivity and conditional exchangeability of the treatment adherence. 
If loss to follow-up is present, then the assumptions of positivity and sequential ignorability for this censoring process are also required for identifying both intention-to-treat and per-protocol effects.
Appendix~\ref{assumptions}  provides details of the causal estimands and assumptions for interested readers, where the formal mathematical language for the potential outcome framework of causal inference is used along with some clarifications using plain language.

\subsubsection{Estimating the intention-to-treat effect} \label{ITTestimation}

We now describe the steps of estimating the intention-to-treat effect implemented in \pkg{TrialEmulation}.  
\begin{enumerate}[label=(\roman*)]
\item \emph{Estimate inverse probability of censoring weights}.  \pkg{TrialEmulation} uses inverse probability of censoring weighting to address dependent censoring (e.g., due to individuals' withdrawal from the trial) \citep{Robins2000,Young2020}.  If dependent censoring is absent in a specific trial, then Step (i) can be omitted.

 At a  follow-up  visit $k$ ($k \geq 1$),
the stabilised inverse probability of censoring  weight for each individual is defined as
\begin{equation}\label{sw_censor}
\mbox{SW}^C_{k}=\prod_{j=0}^{k-1} \frac{\
\mbox{Pr}(C_{j} =0 \mid {C}_{j-1}=Y_j=0, {A}_{0},  V, L_0)}{\mbox{Pr}(C_{j}=0 \mid {C}_{j-1}=Y_j=0, \overline{A}_{j},  V,  \overline{L}_{j})}.
\end{equation}
Note that $\mbox{SW}^C_{0}=1$ for $k=0$, so inverse probability of censoring weighting is not applied at baseline.
 The corresponding unstabilised weight is given by the same formula but with the numerator of equation~(\ref{sw_censor}) replaced by 1. Stabilisation can alleviate the problem of some individuals having a probability
of being uncensored close to $0$, which causes their unstabilised weights
to be very large.
Note that the covariates ($V$ and $L_0$) that are conditioned upon in the numerator terms of the stabilised weights must be included as covariates in the marginal structural model described in Step (ii) below.

To estimate the terms in the denominator of equation~\eqref{sw_censor}, we often specify a  logistic model, for example,
\begin{equation}
   \mbox{logit}\{\mbox{Pr}(C_{j}=0 \mid  {C}_{j-1}=Y_j=0, \overline{A}_{j-1}, A_j=a,V,  \overline{L}_{j})\}=\alpha_{0,a}+{\alpha}_{1,a}\trans V+{\alpha}_{2,a}\trans L_j, \nonumber
\end{equation}
 for $a=0,1$ and $j=0, \ldots, K-1$ (by convention $C_{-1}=0$), where $\alpha_{0,a}$, ${\alpha}_{1,a}$ and ${\alpha}_{2,a}$ are regression parameters. This model assumes that the conditional probability of being uncensored depends only on the most recently received treatment, the baseline covariates and the most recent values of time-varying covariates.
A
correct specification of this model is required to achieve consistent
estimation of the intention-to-treat effect. Therefore,
care needs to be taken when choosing this model.

To estimate the terms in the numerator of equation~\eqref{sw_censor}, a logistic model can also be specified, for example,
$$
\mbox{logit}\{\mbox{Pr}(C_{j}=0 \mid  {C}_{j-1}=Y_j=0, A_0=a, V,  {L}_{0})\}=\theta_{0,a}+{\theta}_{1,a}\trans V+{\theta}_{2,a}\trans L_0,
$$  for $a=0,1$,  where $\theta_{0,a}$, ${\theta}_{1,a}$ and ${\theta}_{2,a}$ are regression parameters.
This model does not need to be correctly specified to achieve consistent
estimation of the intention-to-treat effect.

\item \emph{Fit a marginal structural model by applying a weighted pooled logistic regression model to the observed outcome data}. By
weighting the observed outcome indicator $Y_k$  for each individual 
with the inverse probability of censoring weights in
a  pooled logistic regression model for the discrete-time hazard, we are able to fit a marginal structural model for the potential time-to-event outcome as a function of the assigned treatment  $A_0$ at baseline,   baseline
covariates $V$ and baseline values of time-varying covariates $L_0$.  This is because under the assumptions made in Appendix~\ref{assumptions},  
$$\mbox{Pr}(Y_{k}^{a} =1 \mid Y_{k-1}^{a}=0, V, L_{0})=\mbox{Pr}({Y}_{k}=1 \mid Y_{k-1}=0, {A}_{0}=a, V, {L}_{0}).
$$

For example, the model
\begin{equation}\label{MSM-ITT}
  \mbox{logit}\{\mbox{Pr}({Y}_{k}=1 \mid Y_{k-1}=0, {A}_{0}=a, V, {L}_{0})\}=\beta_{0}+\beta_1k + \beta_2k^2 +\beta_3a+{\beta}_4\trans V+ {\beta}_5\trans L_0
\end{equation}
 allows the baseline hazard to vary by visit $k$ ($k=0, \ldots, K$) with a quadratic functional form (by convention $Y_{-1}=0$). Splines can also be included for flexible modelling of the baseline hazard function.
 Interactions between $A_0$ and $V$ and/or $L_0$ can also be included.

\item \emph{Create two data sets with all individuals assigned to be treated and untreated, respectively}. In preparation for estimating the marginal intention-to-treat effect, we
create two data sets with a copy of the covariates ($V$ and $L_0$) from every individual
at the baseline of the emulated trial.  The data sets include a variable indicating
that all individuals are assigned to the treatment or non-treatment (i.e., setting $A_0 = 1$ for one data set and
$A_0 = 0$ for the other data set).

\item \emph{Predict conditional cumulative incidences for each individual
in both data sets}. 
Let $h^a(k)=\mbox{Pr}(Y_{k}^{a} = 1 \mid Y_{k-1}^{a}=0, V, L_{0})$. The conditional cumulative incidence $\mbox{Pr}(Y_{k}^{a} =1 \mid  V, L_{0})$ can be calculated as 
$$\sum_{k=0}^K \left[h^a(k) \prod_{j=0}^{k-1} \{1-h^a(j)\}\right], $$ where $h^a(k)$  can be predicted using the fitted marginal structural model in Step (ii).

\item \emph{Estimate marginal cumulative incidences for treated and untreated and calculate their differences}.  
Since 
$$
 \mbox{Pr}(Y_{k}^{a} =1  )= {\rm E}_{V, L_{0}}\{\mbox{Pr}(Y_{k}^{a} =1 \mid  V, L_{0})\},  
$$ we average the predicted conditional cumulative incidences in Step (iv) at each visit over all individuals in the two data sets separately and then take the difference to estimate $\mbox{Pr}(Y_{k}^{a} =1)-\mbox{Pr}(Y_{k}^{a} =1)$ defined in~\eqref{ITTestimand}.

\end{enumerate}

\subsubsection{Estimating the per-protocol effect} \label{PPestimation}

We now describe the steps of estimating the per-protocol effect implemented in \pkg{TrialEmulation}. Note that  Steps (f)-(h) are identical to Steps (iii)-(v) in Section~\ref{ITTestimation} for estimating the marginal intention-to-treat effect.

 \begin{enumerate}[label=(\alph*)]
  \item  \emph{Artificially censor individuals' follow-up when they stop adhering to the treatment assigned at the baseline of the emulated trial} \citep{Danaei2013,Murray2021}.

 \item \emph{Estimate inverse probability of treatment weights}.
 We need to address the selection bias from the artificial censoring in Step (a)  as it is induced by non-adherence affected by baseline and time-varying covariates, which are also associated with the time-to-event outcome.
\pkg{TrialEmulation} uses inverse probability of treatment  weighting to address the selection bias due to non-adherence.

At a given follow-up visit $k$ ($k \ge 1$), the stabilised inverse probability of treatment  weight for each individual is  defined as
\begin{equation}\label{IPTW}
\mbox{SW}^A_{k}=\prod_{j=1}^k \frac{\
\mbox{Pr}(A_{j} =a \mid Y_{j-1}=C_{j-1}=0, \overline{A}_{j-1}=(a,\ldots,a), V, L_0 )}{\mbox{Pr}(A_{j}=a \mid Y_{j-1}=C_{j-1}=0, \overline{A}_{j-1}=(a,\ldots,a),  V, \overline{L}_j)},
\end{equation}
 for $a=0, 1$. Note that $\mbox{SW}^A_{0}=1$ so inverse probability of treatment weighting is not applied at baseline. The unstabilised version of~\eqref{IPTW} would replace the numerator terms with 1. The benefit of stabilisation of inverse probability of treatment weights is similar to that for inverse probability of censoring weights. Again, the covariates included in the numerator terms of~\eqref{IPTW}    must also be included in the marginal structural model described in Step (c) below.

The denominator terms in~\eqref{IPTW} are often estimated by specifying a logistic model, and a correctly specified model is required for consistent estimation of the per-protocol effect.  For example, we can specify, separately for $a=0, 1$,
\begin{eqnarray}
 &&\mbox{logit}\{\mbox{Pr}(A_{j}=1 \mid   Y_{j-1}=C_{j-1}=0, \overline{A}_{j-1}=(a,\ldots,a), V,  \overline{L}_{j})\} \nonumber\\&=&\gamma_{0,{a}}+\gamma_{1,{a}}(j-1)+\gamma_{2,{a}}(j-1)^2+ \gamma_{3,{a}}\trans V+\gamma_{4,{a}}\trans L_j,  \nonumber
\end{eqnarray}
 where $j=1, \ldots, K$. In this model, the conditional probability of adhering to the assigned treatment depends on the time passed since assignment (i.e.,  $j-1$, also called `time on the treatment regime'), baseline covariates and recent values of time-varying covariates.

Similarly, to estimate the numerator terms in~\eqref{IPTW},
a logistic model can be used. For example, we can specify, separately for $a=0, 1$,
\begin{eqnarray}
    &&\mbox{logit}\{\mbox{Pr}(A_{j}=1 \mid  Y_{j-1}=C_{j-1}=0,  \overline{A}_{j-1}=(a,\ldots,a), V,  L_0)\} \nonumber\\
    &=&\eta_{0,{a}}+\eta_{1,{a}}(j-1)+\eta_{2,{a}}(j-1)^2+ \eta_{3,{a}}\trans V+\eta_{4,{a}}\trans L_0, \nonumber
\end{eqnarray}
 where $j=1, \ldots, K$. A misspecification of the model for the numerator terms in~\eqref{IPTW} would lead to inefficient
or unstable inverse probability of treatment weights but it does not affect the consistent estimation of the per-protocol effect, provided that the model for the
denominator terms of~\eqref{IPTW} is correctly specified. It is good practice to check the mean of estimated stabilised weights, since a necessary condition for the correct model specification for the denominator terms is that the stabilised weights have a mean of one \citep{Cole2008}.

\item \emph{Estimate inverse probability of censoring weights}. This step is identical to Step (i) in Section~\ref{ITTestimation} except that the data for estimating the inverse probability censoring weights only include observations after applying the artificial censoring in Step a) above.

\item \emph{Calculate the product of the estimated inverse probability of treatment and censoring weights $\widehat{SW}_{k}=\widehat{\mbox{SW}}^A_{k}\widehat{\mbox{SW}}^C_{k}$}.

In practice, the weights $\widehat{SW}_{k}$ can be highly variable when many covariates are included in the models for estimating the weights.  In this situation, users may wish to truncate the weights (e.g., at the 99th percentile); the \pkg{TrialEmulation} package includes an option to implement this.

\item \emph{Fit a marginal structural model by applying a weighted pooled logistic regression model to the observed outcome data}. By
weighting the observed outcome indicator $Y_k $  with $\widehat{SW}_{k}$ in
a  pooled logistic regression model for the discrete-time hazard, we can 
fit a marginal structural model for the potential time-to-event outcome as a function of the assigned treatment and non-treatment, the follow-up visit,   baseline
covariates $V$ and baseline values of time-varying covariates $L_0$.
This is because under the assumptions made in Appendix~\ref{assumptions},  
$$\mbox{Pr}(Y_{k}^{\overline{a}_k} =1 \mid Y_{k-1}^{\overline{a}_{k-1}}=0, V, L_{0})=\mbox{Pr}({Y}_{k}=1 \mid Y_{k-1}=0, \overline{A}_{k}={\overline{a}_k}, V, {L}_{0}),
$$ for $\overline{a}_k =  (0, \ldots, 0)$ and  $\overline{a}_k =  (1, \ldots, 1)$.

For example, we can specify
the following model
$$
\mbox{logit}\{\mbox{Pr}({Y}_{k}=1 \mid Y_{k-1}=0, \overline{A}_{k}=\overline{a}_{k}, V, {L}_{0})\}=\beta_{0}+\beta_1k + \beta_2k^2 +\beta_3(k) a+\beta_4\trans V+ \beta_5\trans L_0,
$$  which allows the hazard ratio $\beta_3(k)$ to vary by $k$ ($k=0, \ldots, K$), i.e.,  time length of being  adherent to treatment. Parametric functional forms can be applied to model $\beta_3(k)$ or  splines can  be included for flexible modelling.

\item \emph{Create two data sets with all individuals assigned to be treated and untreated, respectively}.
\item \emph{Predict conditional cumulative incidences for each individual
in both data sets}. 

\item \emph{Estimate marginal cumulative incidences for treated and untreated and calculate their differences}.  

\end{enumerate}

\subsubsection{Constructing confidence intervals of treatment effects} \label{CImethods}
We use the sandwich variance estimator to estimate robust standard errors of the parameters in the pooled logistic regression models, assuming the estimated weights are fixed \citep{Hernan2000,Danaei2013}. 
Specifically, the sandwich variance estimator is
$$
   \left\{\sum_{i=1}^n \frac{\partial \bm{U}_i(\hat{\bm{\beta}})}{\partial \bm{\beta}^{\text{T}}}\right\}^{-1} \left\{ \sum_{i=1}^n \bm{U}_i(\hat{\bm{\beta}}) \bm{U}_i(\hat{\bm{\beta}})^{\text{T}}\right\} \left\{\sum_{i=1}^n \frac{\partial \bm{U}_i(\hat{\bm{\beta}})}{\partial \bm{\beta}^{\text{T}}}\right\}^{-1},
$$
where $\hat{\bm{\beta}}$ are estimates of the parameters $\bm{\beta}$ in the pooled logistic regression models, and   $\bm{U}_i(\hat{\bm{\beta}})$ is the weighted score function of the pooled logistic models evaluated at $\hat{\bm{\beta}}$.

To obtain confidence intervals for marginal intention-to-treat and per-protocol effects, we follow the approach of \cite{Mandel2013} and construct  simulation-based confidence intervals as follows:

\begin{enumerate}
\item[1)] Draw a sample  $\bm{\beta}_s$ ($s=1, \ldots S$) from a multivariate Normal distribution $\mbox{MVN}( \hat{\bm{\beta}}, \hat{\bm{\Sigma}})$, where $\hat{\bm{\Sigma}}$ is the sandwich variance estimate of $\hat{\bm{\beta}}$.

\item[2)] For each $s = 1, ..., S$, estimate the marginal intention-to-treat and per-protocol effects 
by setting the coefficients in the pooled logistic models equal to $\bm{\beta}_s$. That is, repeat Steps (iv)-(v) in Section~\ref{ITTestimation} and Steps (g)-(h) in Section~\ref{PPestimation} for each sample $\bm{\beta}_s$.

\item[3)] Use the 2.5\% and 97.5\% percentiles of $S$ cumulative incidence difference estimates 
at each follow-up visit as estimated lower and upper-bounds of the 95\%
confidence interval of cumulative incidence difference at that visit.

\end{enumerate}

\subsection{Emulating a sequence of target trials}\label{sequencetrial}
After introducing the notations, assumptions, causal estimands and analysis steps in a single emulated trial, we are now ready to describe the methodology of emulating a sequence of target trials.  In Section~\ref{intro}, we have briefly explained the rationale behind emulating a sequence of trials, and now to help understand the main idea,  we will start with a schematic illustration.  Then we will introduce additional notation to describe the analysis steps. The causal assumptions remain the same as in Section~\ref{singletrial} and the causal estimands are similarly defined.

\subsubsection{Illustration of the settings}

 \begin{figure}[t!]
\centering
\includegraphics[width=\textwidth]{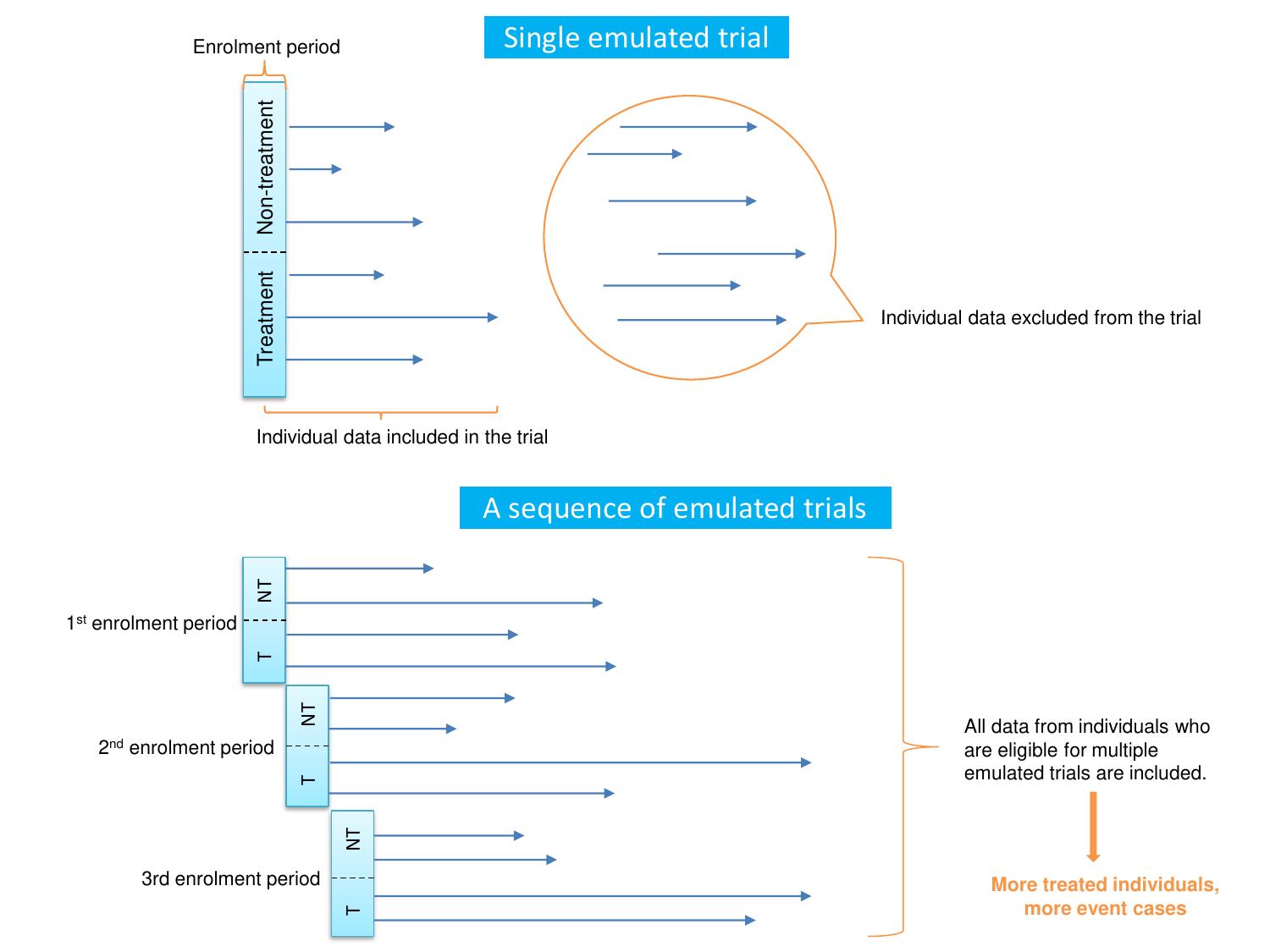}
\caption{\label{illustration} A schematic illustration of a single emulated trial and a sequence of emulated trials. Treatment:  group of eligible individuals who started to receive the treatment in the trial enrolment period; Non-treatment:  group of individuals who have not received the treatment in the trial enrolment period; T: treatment; NT: non-treatment.}
\end{figure}

Figure~\ref{illustration} provides a schematic illustration of a sequence of emulated trials in contrast with a single emulated trial. For a single emulated trial,  an enrolment period (e.g., during a calendar month) is usually specified and all individuals who satisfied the eligibility criteria in this enrolment period would be included in this emulated trial. However, for those individuals who entered the observational cohort/database later than the enrolment period but otherwise satisfied the trial eligibility criteria, their data would be excluded from analyses of the single emulated trial. Moreover, for eligible individuals who had not been on treatment at the enrolment period but later started to receive the treatment, their follow-up would be artificially censored for the per-protocol analysis of the emulated trial, meaning that the data after they started the treatment are discarded. As a result, the number of eligible individuals who were in the treatment group and the number of events that occurred among them can be very small.

The main idea of emulating a sequence of target trials is to utilise available information from the observational cohort/database more efficiently by allowing multiple enrolment periods with the same trial eligibility criteria. Using the hypothetical target trial example of the antiglycaemic drug described in Table~\ref{protocol}, suppose that January of a calendar year was the enrolment period for the first trial.  In this first trial, eligible individuals were followed up for up to 5 years (60 months).  For the second trial, the enrolment period could be, e.g., February of the same calendar year.  In this second trial, some individuals who were included in the non-treatment group of the first trial may be included in the treatment group of the second trial if they started to receive the antiglycaemic treatment during the second enrolment period and also satisfied the trial eligibility criteria.  The enrolment period of this second trial defines its baseline, and the maximum follow-up length in this trial is 59 months. If we take each month of this calendar year as the enrolment period for a separate trial, then we can emulate 12 trials, each with the same eligibility criteria and treatment strategies for comparison but with different lengths of maximum follow-up (49-60 months). For the analysis, the data from the original observational cohort/database are \emph{expanded} to create a larger data set that contains the data for each of the sequence of the emulated trials.   This expanded data set can be analysed altogether for a more efficient estimation of causal treatment effects.

\subsubsection{Additional notation}

Now let us  introduce the additional notation for a sequence of emulated trials, which largely follows
the notation used in \cite{Danaei2013}.
Consider the hypothetical example of 12 emulated trials described in the last section. We use $m = 0, ..., 11$ to index the emulated
trial, and $k$ ($k=0, 1, 2, \ldots$) to refer to the follow-up month in a trial relative to the baseline time of that trial.
The maximum number of follow-up months in trial $m$ is denoted $K_m$.
For example, an individual enrolled in trial $0$ has baseline and possible follow-up months $k = 0, 1, ..., 60$ (i.e.,\ $K_m=60$); an individual enrolled in trial $1$  has baseline and possible follow-up months $k = 0, 1, ..., 59$, and so on.
For each individual in the observational database, let $E_m$ be the binary indicator of whether that individual was eligible ($E_m = 1$) or not ($E_m = 0$) for trial $m$.
If $E_m=1$, let $Y_{m,k}$ be the binary variable indicating whether this individual experienced the event of interest by the end of follow-up month $k$ in trial $m$.
Similarly, if $E_m=1$, let $A_{m,k}$ and $L_{m,k}$ denote the individual's treatment and time-varying covariates at follow-up month $k$ in trial $m$, and let $C_{m,k}$ be a binary indicator that the individual was censored between the follow-up months $k$ and $k+1$ in trial $m$.
If $E_m=0$, define $Y_{m,k} = A_{m,k} = L_{m,k} = C_{m,k} = 0$.

\subsubsection{Causal estimands and assumptions}
The causal estimands in a sequence of emulated trials are similarly defined as in a single emulated trial. 
The set of marginal structural models fitted by pooled logistic regressions, one for each trial, can be themselves pooled across trials, with some functional form(s) of the trial index $m$ appearing as covariate(s) in this single pooled model \citep{Danaei2013}.
The marginal intention-to-treat and per-protocol effects need to be interpretable to a target trial population that is characterised by some baseline covariate distributions. In our hypothetical example, the target trial population could be the group of individuals who are eligible for a particular one of the 12 trials or alternatively a specific population to which the antiglycaemic treatment strategy is intended to be applied.

The causal assumptions stated in the single trial setting are also required for
causal effect identification and estimation in the sequential trials setting, with the only difference that these
assumptions apply to all variables in all emulated trials.

\subsubsection{Analysis}
There are a few subtle differences when performing intention-to-treat and per-protocol analyses for a sequence of emulated trials compared with a single emulated trial, for which we will provide details below. Otherwise, the analysis steps are exactly the same as in the single trial setting.

\begin{enumerate}
\item[1)] \emph{Estimate the inverse probability of  censoring and treatment weights using  the original observational data}.

The stabilised inverse probability of censoring  weight for each individual in trial $m$  at a given follow-up month $k$ ($k \ge  1$)  is defined as
\begin{equation}\label{sw_censor_m}
\mbox{SW}^C_{m,k}=\prod_{j=0}^{k-1} \frac{\
\mbox{Pr}(C_{m,j} =0 \mid {C}_{m,j-1}=Y_{m,j}=0, {A}_{m,0},  V, L_{m,0}, E_m=1)}{\mbox{Pr}(C_{m,j}=0 \mid {C}_{m,j-1}=Y_{m,j}=0, \overline{A}_{m,j},  V,  \overline{L}_{m,j}, E_m=1)}.
\end{equation} These weights will be applied to outcome observations in the emulated trials.
Following \cite{Danaei2013}, the probabilities in the numerator and denominator of equation~(\ref{sw_censor_m}) are estimated by fitting logistic regression models to the original observational data, rather than to the expanded data set.
Specifically, we use the observed censoring indicators of each individual from the follow-up months that correspond to the baselines of the eligible trials until the last trial follow-up months that could be included in the intention-to-treat analysis or the per-protocol analysis. Duplicates of the censoring indicators within individuals (if eligible for multiple trials) are removed.

For the per-protocol analysis, the stabilised inverse probability of treatment weight (to deal with artificial censoring due to treatment switching) for each individual in trial $m$  at follow-up month $k$ ($k \ge 1$)  is defined as
\begin{equation}\label{IPTW_m}
\mbox{SW}^A_{m,k}=\prod_{j=1}^k \frac{\
\mbox{Pr}(A_{m,j} =a \mid  Y_{m,j-1}=C_{m,j-1}=0,\overline{A}_{m,j-1}=(a, \ldots, a), V, L_{m,0}, E_m=1)}{\mbox{Pr}(A_{m,j}=a \mid  Y_{m,j-1}=C_{m,j-1}=0, \overline{A}_{m,j-1}=(a, \ldots, a),  V, \overline{L}_{m,j},E_m=1)},
\end{equation}
 for $a=0, 1$. Just as with equation~\eqref{sw_censor_m}, the logistic regression models for the numerator and denominator terms in~\eqref{IPTW_m} are fitted to the original observational data.  Following \cite{Danaei2013}, we use the observed treatment indicators of each individual from the follow-up months that correspond to the baselines of the eligible trials until the trial follow-up months where the individuals stopped adhering to the assigned treatments or the last trial follow-up months. Duplicates of the treatment indicators within individuals (if eligible for multiple trials) are removed.

For intention-to-treat analyses, only the estimated inverse probability of censoring weights,  $\widehat{\mbox{SW}}^C_{m,k}$, are required. For per-protocol analyses, we calculate the product of the estimated inverse probability of treatment and censoring weights $\widehat{\mbox{SW}}_{m,k}=\widehat{\mbox{SW}}^A_{m,k}\widehat{\mbox{SW}}^C_{m,k}$.

\item[2)] \emph{Specify the marginal structural model for combining analyses of multiple emulated trials}.

With the estimated weights, we can fit a marginal structural model by applying a weighted pooled logistic model for the discrete-time hazard to all observed outcome indicators from all emulated trials. Different specifications of the pooled logistic model reflect assumptions about how the model parameters could vary by trials. For example, the model
\begin{eqnarray}
&&\mbox{logit}\{\mbox{Pr}({Y}_{m,k}=1 \mid Y_{m,k-1}=0, {A}_{m, 0}=a, V, {L}_{m,0}, E_m=1)\} \nonumber\\
&=&\beta_{0} (m) +\beta_1k + \beta_2k^2 +\beta_3a+\beta_4\trans V+ \beta_5\trans L_{m,0}
\end{eqnarray}
differs from the model in~\eqref{MSM-ITT} by including a trial-specific intercept $\beta_{0} (m)$, which can be modelled by parametric forms or splines. In addition, this model assumes that the hazard ratio of treatment assignment,  $\beta_3$, does not vary by trials.  Therefore information can be borrowed across the trials to estimate $\beta_3$ more efficiently. If we do allow the hazard ratio of treatment assignment to vary by trials, we could introduce some parametric form $\beta_3(m)$ and test this interaction between treatment assignment and trials.  If this interaction is significant, it may suggest some treatment effect heterogeneity across calendar time, possibly due to changes in clinical practice.  

\item[3)] \emph{Define the target trial population when estimating marginal treatment effects}.
It is important to choose a target trial population, that is, a distribution of baseline covariates, as well as a value of trial index $m$, as both of these determine the interpretation of the marginal intention-to-treat and per-protocol effects.
In practice, one might define the target trial population to be the population of individuals who were eligible for the most recent emulated trial (and take $m$ to be the index of that trial), or to be some other population to which the causal treatment effect is most relevant \citep{Keogh2021}.
\end{enumerate}

\section{Overview of the package}\label{pkgdetails}

\pkg{TrialEmulation} provides a set of flexible functions that can be run on the observational data provided by the user, in order to create an expanded data set corresponding to the sequence of target trials and to perform intention-to-treat and per-protocol analyses on this expanded data set.

\subsection{Data  processing}\label{data1}

The function \fct{data\_preparation}  expands the user-provided observational data to emulate a sequence of target trials and also estimates the inverse probability of treatment and censoring weights as required.
The observational data provided by users should be in the person-time format, i.e., the  `long' format.
Specifically, the required variables for each person-time observation are the following: (1) an identifier number for the individuals; (2) the index of the visit/period; (3) the treatment indicator for that visit/period; (4) the indicator of the outcome event at that visit/period; (5) the indicator of eligibility for the target trial (using whatever criteria defined by users) at that visit/period; (6) any time-invariant and time-varying covariates that could confound the effect of treatment on the risk of the outcome event.  If dependent censoring needs to be addressed, then an indicator variable for the censoring status is required, with 1 being censored and 0 being uncensored, along with any covariates that affect the censoring process.

The most important arguments of \fct{data\_preparation} are
\begin{Code}
data_preparation(data,
  id = "id", period = "period", treatment = "treatment",
  outcome = "outcome", eligible = "eligible", 
  outcome_cov = ~ 1, model_var = NULL,
  estimand_type = c("ITT", "PP", "As-Treated"),
  switch_d_cov = ~ 1, switch_n_cov = ~ 1,
  use_censor_weights = FALSE, cense = NA, 
  pool_cense = c("none", "both", "numerator"),
  cense_d_cov = ~ 1, cense_n_cov = ~ 1,
  data_dir,  save_weight_models = FALSE,
  glm_function = "glm",
  chunk_size = 500, separate_files = FALSE, ...),
\end{Code}
where \code{data} is the observational data in the person-time format saved in a \class{data.frame}, and the arguments in the second and third lines (\code{id}, \code{period}, \code{treatment}, \code{outcome}, \code{eligible}) specify the variable names in \code{data}  that correspond to the individual identifier number, the visit/period number, the treatment indicator, the outcome indicator and the eligibility indicator, as described above. 

\code{outcome_cov} is the formula to specify the baseline covariates to be included in the marginal structural  model for emulated trials. Note that if a time-varying covariate is specified in \code{outcome_cov}, only its value at each of the trial baselines will be included in the expanded data.  \code{model_var} specifies the treatment variables to be included in the expanded data.  \pkg{TrialEmulation} can create either of two treatment variables.
\code{model_var = "assigned_treatment"} will create a variable \code{assigned_treatment} that is the assigned treatment at the trial baseline, while \code{model_var = "dose"} will create a variable \code{dose} that is the cumulative number of  treatments received
since the trial baseline.  The variable \code{dose} might be of interest in a per-protocol analysis or for fitting standard marginal structural models for all possible treatment sequences (see Appendix D). 
If \code{model_var = c("assigned_treatment", "dose")} is specified, then both these treatment variables will be included in the expanded data returned by \fct{data\_preparation}.

\code{estimand_type} is the option to specify the estimands for the causal analyses in emulated trials. \code{estimand_type = "ITT"} will prepare data for intention-to-treat analyses, where treatment switching after trial baselines are ignored.  \code{estimand_type = "PP"} will prepare data for per-protocol analyses, where individuals' follow-ups are artificially censored and inverse probability of treatment weighting is applied. \code{estimand_type = "As-Treated"} will prepare data for fitting standard marginal structural model for all possible treatment sequences after trial baselines, where individuals' follow-ups in emulated trials are \emph{not} artificially censored but treatment switching after trial baselines are accounted for by applying inverse probability of treatment weighting \citep{Robins2000,Hernan2000}. This is similar to fitting standard marginal structural models using observational cohort data, except that in target trial emulation we introduce multiple trial enrolment periods, therefore different marginal structural models can be specified for different trials.   \pkg{TrialEmulation} can also be used to fit standard marginal structural models using observational cohort data; see more details in Appendix~\ref{fitmsm}.
Note that we follow the SAS macro  \pkg{initiators} \citep{Logan2011} to use the argument value \code{`As-Treated'} for the option of fitting marginal structural models for all possible treatment sequences in emulated trials. If this option is specified, only the parameter estimates of fitted marginal structural models are provided,   but not marginal treatment effects based on cumulative incidences.

\code{switch_d_cov} and \code{switch_n_cov} are the formulas to specify the logistic models for estimating the denominator and numerator terms in~\eqref{IPTW_m} for the inverse probability of treatment weights, respectively. If unstabilised weights are required, \code{switch_n_cov} should be omitted. 
A derived variable named \code{time\_on\_regime} containing the duration of time that the individual has been on the current treatment/non-treatment is available for use in these models.

\code{use_censor_weights} is the option to require the inverse probability of censoring weights. If \code{use_censor_weights = TRUE}, then the variable name of the censoring indicator needs to be provided in the argument \code{cense}.
\code{cense_d_cov} and \code{cense_n_cov} are the formulas to specify the logistic models for estimating the denominator and numerator terms in~\eqref{sw_censor_m} for the inverse probability of censoring weights, respectively. 
If unstabilised weights are required, \code{cense_n_cov} should be omitted. \code{pool\_cense} is the option to specify whether two separate censoring models are fitted for those treated and those untreated at the immediately previous visit. When \code{pool\_cense = "both"}, a pooled censoring model for those treated and those untreated at the immediately previous visit is fitted for both the denominator and numerator terms in~\eqref{sw_censor_m}. Note that if \code{estimand_type = "ITT"}, then a pooled censoring model for those treated and those untreated at the immediately previous visit \emph{must} be specified for the \emph{numerator} terms in~\eqref{sw_censor_m}. Therefore, \code{estimand_type = "ITT"} would only allow \code{pool\_cense = "both"} or \code{pool\_cense = "numerator"}, but forbid \code{pool\_cense = "none"}, where two separate censoring models are fitted for those treated and those untreated at the immediately previous visit for both numerator and denominator terms in~\eqref{sw_censor_m}. 
There is no restrictions for options in \code{pool\_cense} if \code{estimand_type = "PP"} or \code{estimand_type = "As-Treated"}.

If \code{save_weight_models = TRUE}, the fitted models for obtaining weights are saved as \code{glm} objects in the directory specified in the argument \code{data_dir}. 
The default function to fit the models for the weights is the \fct{glm} function in the \pkg{stats} package. Users can also specify   \code{glm\_function = "parglm"} such that the \fct{parglm} function in the \pkg{parglm} package \citep{Christoffersen2022} can be used for fitting generalized linear models in parallel. The default control setting for  \fct{parglm} is \code{nthreads = 4} and \code{method = "FAST"}, where four cores and Fisher information are used for faster computation. Users can change the default control setting by passing the arguments \code{nthreads} and \code{method} in the \fct{parglm.control} function of the \pkg{parglm} package to \fct{data\_preparation}, or alternatively, by passing a \code{control} argument with a list produced by \code{parglm.control(nthreads = , method = )}.

In the case of handling large data sets that exceed the memory limit in \proglang{R} after data expansion, two additional arguments   \code{chunk_size} and \code{separate_files}  can be specified. When \code{separate_files = TRUE}, the user-provided data are processed in chunks of individuals (with size specified in \code{chunk\_size})  and saved into separate CSV files for  each emulated trial. Users can inspect these CSV files in the directory specified in the argument  \code{data\_dir} and manipulate them
easily.

By default,  \fct{data\_preparation} returns the expanded data  in a list that also includes summaries of these data and the fitted models for obtaining the weights. The variable names in the expanded data are self-explanatory, which include \code{id}, \code{trial\_period} (i.e., the trial index), \code{followup\_time} (i.e., the follow-up visit number within trials), \code{outcome}, \code{weight}, \code{treatment}, and derived treatment variables specified in \code{model_var}. The variable \code{weight} contains the estimated weights depending on the estimand type specified and whether inverse probability of censoring weights is required. If \code{estimand_type = "ITT"} and \code{use_censor_weights = FALSE}, then the weights in the variable \code{weight} are equal to one. 
A summary of the results from \fct{data\_preparation}, which includes the fitted models for obtaining the weights,  can be obtained by applying the \fct{summary} function.

\subsection{Case-control sampling for `big data'} \label{sampling}
In practice, emulating trials using data from EHR databases is computationally challenging because data expansion would make   `big data' that cannot be handled within the memory limit of  \proglang{R}.  Moreover, in many situations, only a small proportion of the individuals would experience the event of interest at a given follow-up visit of each of the emulated trials.
We refer to these individuals as the `cases' (at the given follow-up visit in that trial) and the remaining individuals, who had not experienced the event, as the `potential controls' (at the given follow-up visit in that trial). By fitting the pooled logistic regression model to only a random sample of the potential controls (and all the cases), the computational load can be reduced without much loss of statistical efficiency, provided that the number of sampled controls remains considerably greater (e.g.,\ ten times greater) than the number of cases \citep[page 303]{Breslow1987}.

\pkg{TrialEmulation} provides an option to perform case-control sampling to reduce the amount of expanded data, where users are invited to specify a control sampling fraction $p$, e.g.,\ $p=0.01$ or $p=0.001$.
Each potential control independently has a probability $p$ of becoming a selected control, and the expanded data now include only the selected controls and \emph{all the cases} for each of the follow-up visits in each emulated trial.
To account for this inclusion of only a fraction $p$ of the potential controls, the selected controls are each weighted by $1/p$ when fitting the pooled logistic regression model (in addition to the inverse probability of censoring weights and, for per-protocol analyses, the inverse probability of treatment weights).

The function \fct{case\_control\_sampling\_trials}  has the following main arguments,
\begin{Code}
case_control_sampling_trials(data_prep, p_control = NULL, ...),
\end{Code}
where  \code{data\_prep} is the returned object from  \fct{data\_preparation} and \code{p_control} is the control sampling probability specified by  users.

\subsection{Model fitting}

The  \fct{trial\_msm}  function fits a marginal structural model based on a pooled logistic regression using the
\fct{parglm} function with multiple cores or the \fct{glm} function. The robust sandwich variance matrix is calculated
using functions in the \pkg{sandwich} package \citep{Zeileis2020}.

The function \fct{trial\_msm} has the following main arguments,
\begin{Code}
trial_msm(data,
  estimand_type = c("ITT", "PP", "As-Treated"),
  outcome_cov = ~ 1, model_var = NULL,
  include_followup_time = ~ followup_time + I(followup_time^2),
  include_trial_period = ~ trial_period + I(trial_period^2),
  glm_function = c("parglm", "glm"),
  use_sample_weights = TRUE, ...),
\end{Code}
where \code{data} is the object returned by the function \fct{data\_preparation} or by the function
\fct{case\_control\_sampling\_trials}. Derived variables, e.g., splines using the  \fct{ns} function in the \pkg{splines} package and  other terms created by  \proglang{R} model formula operators 
can be added to the formula in \code{outcome_cov}. In addition, \proglang{R} model formulae 
involving derived treatment variables (\code{assigned\_treatment} or \code{dose}) specified in the argument \code{model_var} can be added to \code{outcome_cov}.   
\code{include\_followup\_time} and \code{include\_trial\_period} specify the functional forms of the follow-up visit and trial index numbers to be added to \code{outcome_cov}.  The default would include both the linear and quadratic terms of both variables. The argument \code{glm_function} specifies the package used to fit the pooled logistic model. If \code{data} is obtained by applying case-control sampling, then the argument \code{use\_sample\_weights = TRUE} is required, so that the sampling weights are used when fitting the pooled logistic model.  \fct{trial\_msm} returns the fitted pooled logistic model and also the sandwich variance matrix of the model parameters. A summary of these results can be obtained by applying the \fct{summary} function to the returned object of \fct{trial\_msm}.

Appendix~\ref{moreoptions} provides details of optional arguments in the functions  \fct{data\_preparation}, \fct{case\_control\_sampling\_trials},  and \fct{trial\_msm}.  When data are not required to be expanded in chunks and no case-control sampling is applied, the function \fct{initiators} provides a wrapper of   \fct{data\_preparation} and  \fct{trial\_msm}.

\subsection{Point estimation and inference for marginal cumulative incidences}

The  \fct{predict} function estimates the marginal cumulative incidences when the target trial population receives either the treatment or non-treatment at baseline (for an intention-to-treat analysis) or either sustained treatment or sustained non-treatment (for a per-protocol analysis). The difference between these cumulative incidences is the estimated causal effect of treatment. Currently, the  \fct{predict} function only provides marginal intention-to-treat and per-protocol effects, therefore it is only valid when setting \code{estimand_type = "ITT"} or \code{estimand_type = "PP"}.  The main arguments of \fct{predict} include:
\begin{Code}
predict(object, newdata, predict_times, conf_int = TRUE, 
  type = c("cum_inc", "survival"), ...).
\end{Code}
The argument \code{object} is the returned object from \fct{trial\_msm} or \fct{initiators}.  Users need to provide the baseline covariate data that characterise the target trial population in the argument \code{newdata}, which must have the same columns and formats of variables as in the fitted marginal structural model specified in \fct{trial\_msm}.
The argument \code{predict\_times} specifies the follow-up visits where the cumulative incidences  are estimated. \code{conf_int = TRUE} will estimate the point-wise 95\% confidence intervals of cumulative incidences under treatment and non-treatment and their differences using the procedure described in Section~\ref{CImethods}.  Users can request either cumulative incidences or survival probabilities to be estimated using the option \code{type = c("cum_inc", "survival")}. Details of optional arguments of \fct{predict} can be found in Appendix~\ref{moreoptions}.



\section{Illustrations} \label{sec:illustrations}

\subsection{Data}
We use simulated data based on the algorithm described in \cite{Young2014} to demonstrate the functionality of \pkg{TrialEmulation}. In particular, the simulated data in the \code{data.frame} called \code{simdata_censored} incorporate both treatment switching and dependent censoring that are ubiquitous in real observational cohort settings.   The detailed data generating mechanism is provided in Appendix~\ref{app:simdata} and the \proglang{R} function to simulate these data are provided in the \proglang{R} replication code for this paper.  

Below are the data from the first few individuals.
\begin{Schunk}
\begin{Sinput}
R> load("simdata.Rdata")
R> print(simdata_censored[simdata_censored$ID <= 4,], digits = 2,
+    row.names = F)
\end{Sinput}
\begin{Soutput}
 ID t A X1    X2 X3    X4 age age_s Y C eligible
  1 0 1  0 -0.35  0  0.96  49  1.17 0 1        1
  2 0 1  1 -1.15  1  1.70  30 -0.42 0 0        1
  2 1 1  1  1.45  1  1.70  31 -0.33 0 0        0
  2 2 1  0  1.27  1  1.70  32 -0.25 0 1        0
  4 0 0  0 -1.01  0 -0.31  53  1.50 0 0        1
  4 1 0  0  0.38  0 -0.31  54  1.58 0 0        1
  4 2 1  1 -0.44  0 -0.31  55  1.67 0 0        1
  4 3 1  1  0.20  0 -0.31  56  1.75 0 0        0
  4 4 1  0 -0.45  0 -0.31  57  1.83 0 0        0
  4 5 1  0  0.24  0 -0.31  58  1.92 0 0        0
  4 6 1  1  0.20  0 -0.31  59  2.00 0 0        0
  4 7 0  0 -0.19  0 -0.31  60  2.08 0 0        0
  4 8 1  1 -0.50  0 -0.31  61  2.17 0 0        0
  4 9 1  0  0.43  0 -0.31  62  2.25 0 0        0
\end{Soutput}
\end{Schunk}
The definitions of the variables are as follows. \verb|ID| is the individual's identifier number. \verb|t| is the visit/period number. \verb|A| is the treatment received corresponding to \verb|t|.
\verb|X1| and \verb|X2| are a binary time-varying confounder and a continuous time-varying confounder, respectively.  \verb|X3| and \verb|X4| are a binary time-invariant confounder and a continuous time-invariant confounder, respectively. \verb|age| is the individual's age corresponding to \verb|t|, and \verb|age_s| is the standardised \verb|age| (see detailed definition in Table~\ref{DGM} in Appendix C).
\verb|Y| is the binary indicator of the outcome event corresponding to \verb|t|.  \verb|C| is the censoring indicator  corresponding to \verb|t|.
 \verb|eligible| is the indicator of eligibility for the target trial corresponding to \verb|t|.

Individuals with \verb|ID| $=1$ and \verb|ID| $=2$ were both eligible at \verb|t| $=0$ and both received the treatment, so they belong to the treatment group in the trial with baseline at \verb|t| $=0$. Their follow-ups were censored after  \verb|t| $=0$ and \verb|t| $=2$, respectively.
The individual with \verb|ID| $=4$ satisfied the eligibility criteria at \verb|t| $=0, 1, 2$. He/she will belong to the non-treatment group in the trials with baseline at \verb|t| $=0, 1$ and to the treatment group in the trial with baseline at \verb|t| $=2$, because he/she started to receive the treatment at \verb|t| $=2$. The maximum number of follow-up visits in these data was 9. The individual with \verb|ID| $=4$ completed the follow-up to \verb|t| $=9$, and thus had complete data for the outcome and covariates.  Depending on the type of estimands required, the expanded data for these individuals are different, as illustrated in the following sections.

\subsection{Intention-to-treat analysis under dependent censoring}\label{ITTillustration}

We first use the \fct{data\_preparation} function to expand the data for an intention-to-treat analysis that accounts for dependent censoring. First we set \code{estimand_type = "ITT"} so no artificial censoring is applied.   \code{X1}, \code{X2}, \code{X3}, \code{X4}, \code{age_s} are included in \code{outcome_cov} as covariates in the marginal structural model for the outcome. Note that because \code{X1}, \code{X2} and \code{age_s} are time-varying, only their values at baseline of each trial will be included in the expanded data. The treatment variable of interest in the intention-to-treat analysis  is the assigned treatment, so we set the argument \code{model_var = "assigned_treatment"}.  Because dependent censoring needs to be addressed, the option \code{use_censor_weights} is set at \code{TRUE} as we need to calculate the inverse probability of censoring weights. In the models for the denominator terms of the censoring weights, we include \code{X1}, \code{X2}, \code{X3}, \code{X4} and \code{age_s} in the argument \code{cense_d_cov}. In the models for the numerator terms of the censoring weights, we include  \code{X3}, \code{X4} in the argument \code{cense_n_cov}.   Setting \code{pool_cense = "numerator"}  would only allow separate models to be fitted for those who did and did not receive treatment at the most recent visit for estimating the denominator terms of the censoring weights.   The models for censoring weights are fitted before expanding the data.
  We set \code{save_weight_models = TRUE} such that the fitted model objects are saved in the directory specified in \code{data_dir}.  We suppress the printing of progress messages by setting \code{quiet = TRUE}.

\begin{Schunk}
\begin{Sinput}
R> library("TrialEmulation")
R> working_dir <- getwd()
R> prep_ITT_data <- data_preparation(
+    data = simdata_censored,
+    id = "ID", period = "t", treatment = "A",
+    outcome = "Y", eligible = "eligible",
+    estimand_type = "ITT",
+    outcome_cov =  ~ X1 + X2 + X3 + X4 + age_s,
+    model_var = "assigned_treatment",
+    use_censor_weights = TRUE, cense = "C", 
+    cense_d_cov = ~ X1 + X2 + X3 + X4 + age_s,
+    cense_n_cov = ~ X3 + X4,
+    pool_cense = "numerator", 
+    data_dir = working_dir, save_weight_models = TRUE,
+    glm_function = "parglm", nthreads = 4, method = "FAST",
+    quiet = TRUE)
\end{Sinput}
\end{Schunk}

The expanded data are saved as the first element   of the list  \code{prep\_ITT\_data} returned by \fct{data\_preparation}.
\begin{Schunk}
\begin{Sinput}
R> str(prep_ITT_data, max.level = 1, width = 70, strict.width = 'wrap')
\end{Sinput}
\begin{Soutput}
List of 6
$ data :Classes 'data.table' and 'data.frame': 9069 obs. of 12
   variables:
..- attr(*, ".internal.selfref")=<externalptr>
$ min_period : int 0
$ max_period : int 9
$ N : int 9069
$ data_template:'data.frame': 0 obs. of 12 variables:
$ censor_models:List of 3
- attr(*, "class")= chr [1:2] "TE_data_prep_dt" "TE_data_prep"
\end{Soutput}
\end{Schunk}
There are a few  columns with default names that save the key variables of the expanded data.  The column $id$ is the individual identifier. The column \code{trial_period} is the trial index number and  the column \code{followup_time} is the follow-up visit within each emulated trial.
The column \code{assigned_treatment} is the assigned treatment at the trial baseline.  The column \code{treatment} is the treatment received at each follow-up visit within the trial and the column \code{outcome} is the outcome indicator. \fct{data\_preparation} calculates the  inverse probability of censoring weights and saves them in the column \code{weight}.   Finally, all covariates specified in the argument \code{outcome_cov} are also included.

We now examine the expanded data using the individual with \verb|ID| $=4$ as an example.
Since this individual was eligible  at \verb|t| $ = 0$ and had complete data up to \verb|t| $ = 9$, the expanded data for this individual in  trial $0$  have \code{trial_period} $= 0$,  \code{followup_time} equals 0 up to 9 and  \code{assigned_treatment} $= 0$. Note that we do not apply artificial censoring in the intention-to-treat analysis, so all follow-up visits are included even though this individual switched treatment several times.  Since this individual was also eligible  at \verb|t| $ = 1$, his/her data from \verb|t| $ = 1$ to \verb|t| $ = 9$ are included in  trial $1$ with \code{trial_period} $= 1$ and correspond to \code{followup_time} $= 0$ up to \code{followup_time} $= 8$. He/she still  had \code{assigned_treatment} $=0$ for this trial. Finally, this individual started treatment at \verb|t| $ = 2$ and was eligible for  trial $2$ with \code{trial_period} $= 2$ . Correspondingly, we have \code{followup_time} $= 0$ up to \code{followup_time} $= 7$. Note that  \code{assigned_treatment} $=1$. The data expansion now finishes for this individual because he/she was not eligible for further trials with \code{trial_period} $> 2$.
\begin{Schunk}
\begin{Sinput}
R> prep_ITT_data$data <- prep_ITT_data$data[order(prep_ITT_data$data$id,
+    prep_ITT_data$data$trial_period, prep_ITT_data$data$followup_time),]
R> print(prep_ITT_data$data[prep_ITT_data$data$id == 4,
+    c("id", "trial_period", "followup_time", "outcome", "weight",
+    "treatment", "assigned_treatment")], digits = 2, row.names = F)
\end{Sinput}
\begin{Soutput}
 id trial_period followup_time outcome weight treatment assigned_treatment
  4            0             0       0   1.00         0                  0
  4            0             1       0   0.95         0                  0
  4            0             2       0   0.87         1                  0
  4            0             3       0   0.77         1                  0
  4            0             4       0   0.69         1                  0
  4            0             5       0   0.62         1                  0
  4            0             6       0   0.55         1                  0
  4            0             7       0   0.49         0                  0
  4            0             8       0   0.44         1                  0
  4            0             9       0   0.39         1                  0
  4            1             0       0   1.00         0                  0
  4            1             1       0   0.91         1                  0
  4            1             2       0   0.81         1                  0
  4            1             3       0   0.72         1                  0
  4            1             4       0   0.65         1                  0
  4            1             5       0   0.58         1                  0
  4            1             6       0   0.51         0                  0
  4            1             7       0   0.46         1                  0
  4            1             8       0   0.41         1                  0
  4            2             0       0   1.00         1                  1
  4            2             1       0   0.89         1                  1
  4            2             2       0   0.79         1                  1
  4            2             3       0   0.71         1                  1
  4            2             4       0   0.63         1                  1
  4            2             5       0   0.56         0                  1
  4            2             6       0   0.51         1                  1
  4            2             7       0   0.45         1                  1
 id trial_period followup_time outcome weight treatment assigned_treatment
\end{Soutput}
\end{Schunk}

The values of the covariates \code{X1}, \code{X2}, \code{X3}, \code{X4}, \code{age_s} at each trial baseline are included in the expanded data, see the following as an example. Note that even though \code{X1}, \code{X2} and \code{age_s} are time-varying covariates in \code{simdata_censored}, only their values at trial baselines are used in the expanded data.
\begin{Schunk}
\begin{Sinput}
R> print(prep_ITT_data$data[prep_ITT_data$data$id == 4
+    & prep_ITT_data$data$trial_period == 0,
+    c("id", "trial_period", "followup_time", "X1", "X2", "X3", "X4", "age_s")],
+    digits = 2, row.names = F)
\end{Sinput}
\begin{Soutput}
 id trial_period followup_time X1 X2 X3    X4 age_s
  4            0             0  0 -1  0 -0.31   1.5
  4            0             1  0 -1  0 -0.31   1.5
  4            0             2  0 -1  0 -0.31   1.5
  4            0             3  0 -1  0 -0.31   1.5
  4            0             4  0 -1  0 -0.31   1.5
  4            0             5  0 -1  0 -0.31   1.5
  4            0             6  0 -1  0 -0.31   1.5
  4            0             7  0 -1  0 -0.31   1.5
  4            0             8  0 -1  0 -0.31   1.5
  4            0             9  0 -1  0 -0.31   1.5
\end{Soutput}
\end{Schunk}

A summary of the results of \fct{data\_preparation} can be obtained using the \fct{summary} function.
\begin{Schunk}
\begin{Sinput}
R> summary(prep_ITT_data, digits = 2)
\end{Sinput}
\begin{Soutput}
Expanded Trial Emulation data

        id trial_period followup_time outcome weight treatment X1    X2 X3
   1:    1            0             0       0   1.00         1  0 -0.35  0
   2:    2            0             0       0   1.00         1  1 -1.15  1
   3:    2            0             1       0   1.09         1  1 -1.15  1
  ---                                                                     
9067: 1000            0             7       0   0.73         1  0 -0.30  0
9068: 1000            0             8       0   0.65         1  0 -0.30  0
9069: 1000            0             9       0   0.58         1  0 -0.30  0
         X4 age_s assigned_treatment
   1:  0.96  1.17                  1
   2:  1.70 -0.42                  1
   3:  1.70 -0.42                  1
  ---                               
9067: -0.10  1.00                  1
9068: -0.10  1.00                  1
9069: -0.10  1.00                  1

Number of observations in expanded data: 9069 
First trial period: 0 
Last trial period: 9 

-------------------------------------------------------- 
Weight models
-------------

Censoring models
----------------

censor_models$cens_d0:
Model for P(cense = 0 | X, previous treatment = 0) for denominator 

        term estimate std.error statistic p.value
 (Intercept)     0.97     0.097      10.0 2.5e-23
          X1     0.40     0.114       3.5 4.4e-04
          X2    -0.56     0.060      -9.3 1.7e-20
          X3     0.26     0.113       2.3 2.1e-02
          X4    -0.30     0.058      -5.2 2.2e-07
       age_s     1.04     0.070      14.9 5.6e-50

-------------------------------------------------------- 
censor_models$cens_d1:
Model for P(cense = 0 | X, previous treatment = 1) for denominator 

        term estimate std.error statistic p.value
 (Intercept)     1.98     0.146      13.5 8.3e-42
          X1     0.63     0.249       2.6 1.1e-02
          X2    -0.48     0.096      -5.0 6.6e-07
          X3     0.11     0.189       0.6 5.5e-01
          X4    -0.14     0.105      -1.3 1.8e-01
       age_s     1.16     0.123       9.4 3.4e-21

-------------------------------------------------------- 
censor_models$cens_pool_n:
Model for P(cense = 0 | X) for numerator 

        term estimate std.error statistic  p.value
 (Intercept)    1.984     0.063     31.71 1.1e-220
          X3    0.088     0.089      0.99  3.2e-01
          X4   -0.079     0.046     -1.72  8.6e-02

-------------------------------------------------------- 
\end{Soutput}
\end{Schunk}

As we estimated the inverse probability of censoring weights and set \code{save_weight_models = TRUE}, the fitted model objects for the censoring process are saved in separate files\\  \code{cense_model_d0.rds}, \code{cense_model_d1.rds}, \code{cense_model_pool_n.rds}.
\begin{Schunk}
\begin{Sinput}
R> list.files(working_dir, pattern = "cense_model")
\end{Sinput}
\begin{Soutput}
[1] "cense_model_d0.rds"     "cense_model_d1.rds"    
[3] "cense_model_n0.rds"     "cense_model_n1.rds"    
[5] "cense_model_pool_n.rds"
\end{Soutput}
\end{Schunk}
For example,  \code{cense_model_d0.rds} and \code{cense_model_d1.rds} are the fitted model objects for obtaining the denominator terms of the censoring weights among individuals who did not and individuals who did receive treatment at the most recent visit, respectively.  Since a pooled logistic model is fitted to estimate the numerator terms, only one model object is saved in \code{cense_model_pool_n.rds}. The summaries of these fitted models are also available in the object returned by \fct{data\_preparation}. \\

Using the \fct{trial\_msm} function, we can now specify the marginal structural model using a weighted pooled logistic regression given the baseline values of \code{X1}, \code{X2}, \code{X3}, \code{X4}, \code{age_s}.
\begin{Schunk}
\begin{Sinput}
R> library("splines")
R> ITT_result <- trial_msm(prep_ITT_data$data,
+    estimand_type = "ITT",
+    outcome_cov =  ~ X1 + X2 + X3 + X4 + age_s,
+    model_var = "assigned_treatment",
+    include_followup_time = ~ ns(followup_time, df = 3),
+    glm_function = "glm",
+    use_sample_weights = FALSE,
+    quiet = TRUE)
\end{Sinput}
\end{Schunk}
By default, the linear and quadratic terms of the trial index are included. We use natural cubic splines with 3 degrees of freedom implemented in the \fct{ns} function in the \pkg{splines} package to model the effect of the follow-up visits.  We have suppressed the printing of progress messages by setting \code{quiet = TRUE}. Otherwise, the printed output   will   present the parameter estimates of the marginal structural model with \emph{naive model-based standard errors} first, and then the following table will provide the parameter estimates  with  the \emph{robust standard errors} from the sandwich variance estimator.
The returned object \code{ITT_result} includes a list of the summaries of the fitted model, the parameter estimates with robust standard errors and the sandwich variance matrix. From the summary below,  we can see that the estimated regression coefficient of assigned treatment is $-0.21$ with  95\% confidence interval $[-0.69,  0.27]$.        
\begin{Schunk}
\begin{Sinput}
R> summary(ITT_result, digits = 1)
\end{Sinput}
\begin{Soutput}
Trial Emulation Outcome Model

Outcome model formula:
outcome ~ assigned_treatment + trial_period + I(trial_period^2) + 
    ns(followup_time, df = 3) + X1 + X2 + X3 + X4 + age_s

Coefficent summary (robust):
                      names estimate robust_se  2.5%  97.5%      z p_value
                (Intercept)   -5.081      0.47 -6.00 -4.160 -10.82  <2e-16
         assigned_treatment   -0.212      0.25 -0.69  0.271  -0.86   0.390
               trial_period    0.083      0.19 -0.29  0.452   0.44   0.658
          I(trial_period^2)   -0.130      0.07 -0.27  0.008  -1.84   0.066
 ns(followup_time, df = 3)1    0.408      0.55 -0.68  1.494   0.74   0.462
 ns(followup_time, df = 3)2    0.020      0.55 -1.06  1.099   0.04   0.971
 ns(followup_time, df = 3)3   -0.007      0.56 -1.10  1.086  -0.01   0.991
                         X1   -0.151      0.24 -0.62  0.320  -0.63   0.531
                         X2    0.122      0.11 -0.09  0.340   1.10   0.270
                         X3    1.070      0.33  0.43  1.707   3.29   0.001
                         X4    0.729      0.18  0.38  1.081   4.05   5e-05
                      age_s    0.553      0.21  0.14  0.967   2.62   0.009

ITT_result$model contains the fitted glm model object.
ITT_result$robust$matrix contains the full robust covariance matrix.
\end{Soutput}
\end{Schunk}

\begin{figure}[t!]
\centering
\includegraphics[width=\textwidth]{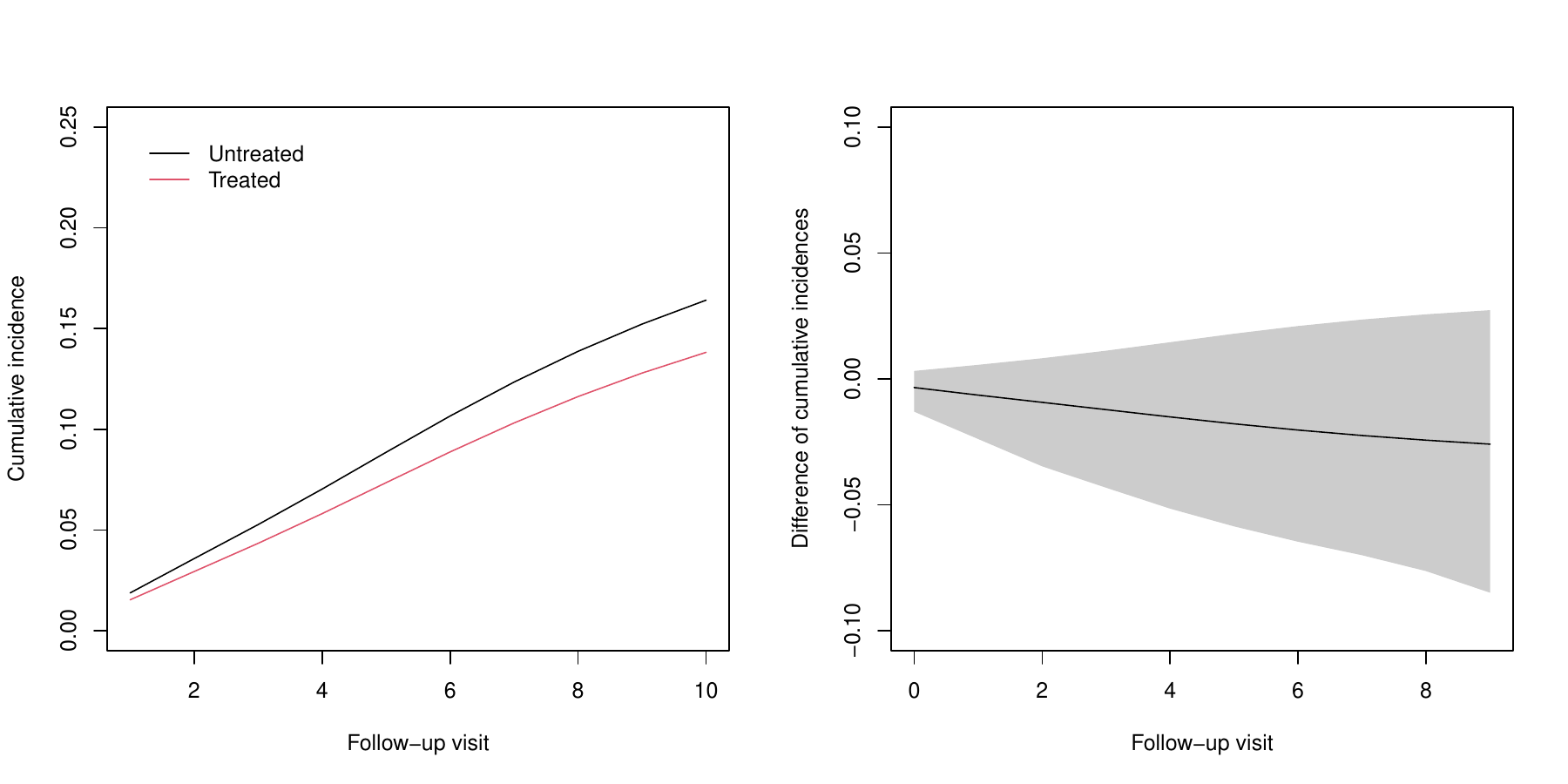}
\caption{\label{fig:ITT} Predicted cumulative incidences for untreated and treated and their differences with point-wise 95\% confidence intervals as the marginal intention-to-treat effect for the trial 0 population. }
\end{figure}

To obtain the marginal intention-to-treat effect in terms of cumulative incidence differences, we need to define the target population to which this effect is applied. For example, below we choose the individuals included in trial 0 as the target population by subsetting the expanded data with \code{trial_period == 0}. To make sure the subsetted data are in the right format (e.g., the factor levels of variables are preserved), we can use the \code{data_template} returned by \fct{data\_preparation}. In the function \fct{predict},  we specify  the fitted marginal structural model returned by \fct{trial\_msm} in the argument \code{object}, and  the follow-up visits for calculating the cumulative incidences in the argument \code{predict_times}.
\begin{Schunk}
\begin{Sinput}
R> new_data <- prep_ITT_data$data[prep_ITT_data$data$trial_period == 0,]
R> new_data <- rbind(data.table::as.data.table(prep_ITT_data$data_template),
+    new_data)
R> set.seed(20222022)
R> ITT_cuminc <- predict(ITT_result, new_data, predict_times = c(0:9))
\end{Sinput}
\end{Schunk}
The returned object from  \fct{predict} is a list of the predicted cumulative incidences and their differences at specified follow-up visits. We can easily plot them to examine the marginal intention-to-treat effect for the trial 0 population as follows in Figure~\ref{fig:ITT}.

\begin{Schunk}
\begin{Sinput}
R> pdf("ITT_cuminc.pdf", width = 12, height = 6)
R> par(mfrow = c(1,2))
R> plot(ITT_cuminc$assigned_treatment_0$cum_inc, ty = "l",
+    ylab = "Cumulative incidence", xlab = "Follow-up visit",
+    ylim = c(0, 0.25))
R> lines(ITT_cuminc$assigned_treatment_1$cum_inc, col = 2)
R> legend(1, 0.25, legend = c('Untreated', 'Treated'), col = 1:2,
+    lty = c(1,1), bty = 'n')
R> followup <- ITT_cuminc$difference$followup_time
R> cum_inc_diff <- ITT_cuminc$difference$cum_inc_diff
R> LL <- ITT_cuminc$difference$`2.5%`
R> UL <- ITT_cuminc$difference$`97.5%`
R> plot(followup, cum_inc_diff, ty = "l",
+    ylab = "Difference of cumulative incidences",
+    xlab = "Follow-up visit", ylim = c(-0.1, 0.1))
R> polygon(c(followup, rev(followup)), c(LL, rev(UL)),
+    col = 'grey80', border = NA)
R> lines(followup, cum_inc_diff)
\end{Sinput}
\end{Schunk}

\subsection{Per-protocol analysis under dependent censoring}
We now use the \fct{data\_preparation} function to expand the data for the per-protocol analysis that accounts for dependent censoring.  We set the option \code{estimand_type = "PP"}, as artificial censoring is required.  The pooled logistic regression for the marginal structural model is the same as in the intention-to-treat analysis.  Therefore the model formula in the \code{outcome_cov} remains the same.   The option \code{use_censor_weights} is set at \code{TRUE} because we need to calculate both the inverse probability of treatment and censoring weights. In the model for the denominator terms of the treatment weights, we include \code{X1}, \code{X2}, \code{X3}, \code{X4} and \code{age_s} in the argument \code{switch_d_cov}. In the model for the numerator terms of the treatment weights, we include  \code{X3}, \code{X4} in the argument \code{switch_n_cov}.
In addition, the linear and quadratic terms of an automatically derived variable \code{time\_on\_regime}, which is defined as the time since treatment initiation,  are included in the treatment model formulae specified in \code{switch_n_cov} and \code{switch_d_cov}.
By default, separate models are fitted for those who did and did not receive treatment at the most recent visit.   These models for treatment weights are fitted to the original data that would be eligible for the per-protocol analysis as described in Section~\ref{sequencetrial}. The options  \code{chunk_size = 500} and  \code{separate_files = TRUE} allow for the data expansion to be done by chunks of 500 individuals and the expanded data by trials are saved as separate CSV  files.  

\begin{Schunk}
\begin{Sinput}
R> prep_PP_data <- data_preparation(
+    data = simdata_censored,
+    id = "ID", period = "t", treatment = "A",
+    outcome = "Y", eligible = "eligible",
+    estimand_type = "PP",
+    outcome_cov =  ~ X1 + X2 + X3 + X4 + age_s,
+    model_var = "assigned_treatment",
+    switch_d_cov = ~ X1 + X2 + X3 + X4 + age_s 
+    + time_on_regime + I(time_on_regime^2),
+    switch_n_cov = ~ X3 + X4 + time_on_regime + I(time_on_regime^2),
+    use_censor_weights = TRUE, cense = "C", 
+    cense_d_cov = ~ X1 + X2 + X3 + X4 + age_s,
+    cense_n_cov = ~ X3 + X4,
+    pool_cense = 'none',
+    data_dir = working_dir, save_weight_models = TRUE,
+    chunk_size = 500, separate_files = TRUE,
+    quiet = TRUE)
\end{Sinput}
\end{Schunk}

We can list these CSV files as follows.
\begin{Schunk}
\begin{Sinput}
R> list.files(working_dir, pattern = "trial_")
\end{Sinput}
\begin{Soutput}
 [1] "trial_0.csv" "trial_1.csv" "trial_2.csv" "trial_3.csv" "trial_4.csv"
 [6] "trial_5.csv" "trial_6.csv" "trial_7.csv" "trial_8.csv" "trial_9.csv"
\end{Soutput}
\end{Schunk}

We can examine the expanded data, using the individual with \verb|ID| $=4$ as an example.
Since this individual start the treatment at \code{t} $=2$, for trial 0, he/she only contributes two observations up to \code{t} $=1$.
\begin{Schunk}
\begin{Sinput}
R> trial_0 <- read.csv("trial_0.csv")
R> print(trial_0[trial_0$id == 4, ], digits = 2, row.names = FALSE)
\end{Sinput}
\begin{Soutput}
 id trial_period followup_time outcome weight treatment X1 X2 X3    X4
  4            0             0       0   1.00         0  0 -1  0 -0.31
  4            0             1       0   0.78         0  0 -1  0 -0.31
 age_s assigned_treatment
   1.5                  0
   1.5                  0
\end{Soutput}
\end{Schunk}
For trial 2, however, this individual now contributes 5 observations from \code{t} $=2$ to \code{t} $=6$ because he/she stopped the treatment at  \code{t} $=7$. These 5 observations correspond to \code{followup_time} $=0, \ldots, 4$ in trial 2.
\begin{Schunk}
\begin{Sinput}
R> trial_2 <- read.csv("trial_2.csv")
R> print(trial_2[trial_2$id == 4, ], digits = 2, row.names = FALSE)
\end{Sinput}
\begin{Soutput}
 id trial_period followup_time outcome weight treatment X1    X2 X3    X4
  4            2             0       0   1.00         1  1 -0.44  0 -0.31
  4            2             1       0   0.84         1  1 -0.44  0 -0.31
  4            2             2       0   1.01         1  1 -0.44  0 -0.31
  4            2             3       0   1.03         1  1 -0.44  0 -0.31
  4            2             4       0   0.86         1  1 -0.44  0 -0.31
 age_s assigned_treatment
   1.7                  1
   1.7                  1
   1.7                  1
   1.7                  1
   1.7                  1
\end{Soutput}
\end{Schunk}

The fitted models for the treatment weights are saved in \code{data_dir}  as separate files.
\begin{Schunk}
\begin{Sinput}
R> list.files(working_dir, pattern = "weight_model_switch")
\end{Sinput}
\begin{Soutput}
[1] "weight_model_switch_d0.rds" "weight_model_switch_d1.rds"
[3] "weight_model_switch_n0.rds" "weight_model_switch_n1.rds"
\end{Soutput}
\end{Schunk}
For example, \code{weight_model_switch_d0.rds} and \code{weight_model_switch_d1.rds} are the fitted model objects for obtaining the denominator terms of the treatment weights among individuals who did not and did receive treatment at the most recent visit, respectively.
We can also examine these fitted models in the summary provided above by \fct{summary}.

In addition to performing data expansion by chunks, we can perform case-control sampling with the \fct{case\_control\_sampling\_trials} function to reduce the amount of data. Note that \fct{case\_control\_sampling\_trials} can be used when \code{separate_files = FALSE} as well.
The amount of data in   \code{simdata_censored} is not large, but we use these data for illustration. Here we sample $50\%$ of the `potential controls' following the method described in Section~\ref{sampling}.
\begin{Schunk}
\begin{Sinput}
R> set.seed(20222023)
R> sampled_data <- case_control_sampling_trials(prep_PP_data, p_control = 0.5)
R> sampled_data <- sampled_data[order(sampled_data$id, sampled_data$trial_period,
+    sampled_data$followup_time),]
R> print(head(sampled_data), digits = 2, row.names = F)
\end{Sinput}
\begin{Soutput}
 id trial_period followup_time outcome weight treatment X1    X2 X3    X4
  1            0             0       0   1.00         1  0 -0.35  0  0.96
  2            0             0       0   1.00         1  1 -1.15  1  1.70
  2            0             1       0   1.06         1  1 -1.15  1  1.70
  4            0             0       0   1.00         0  0 -1.01  0 -0.31
  4            0             1       0   0.78         0  0 -1.01  0 -0.31
  4            1             0       0   1.00         0  0  0.38  0 -0.31
 age_s assigned_treatment sample_weight
  1.17                  1             2
 -0.42                  1             2
 -0.42                  1             2
  1.50                  0             2
  1.50                  0             2
  1.58                  0             2
\end{Soutput}
\end{Schunk}
The returned object from the \fct{case\_control\_sampling\_trials} now includes a new column \code{sample\_weight} for the  case-control sampling weights. In our case, all sampled controls have a weight of two.

Again, we specify the marginal structural model in the \fct{trial\_msm} function.  As we are using the expanded data after case-control sampling, the option \code{use_sample_weights = TRUE} needs to be specified so that the correct case-control sampling weights are applied along with the inverse probability of treatment and censoring weights.
\begin{Schunk}
\begin{Sinput}
R> PP_result <- trial_msm(sampled_data,
+    estimand_type = "PP",
+    outcome_cov =  ~  X1 + X2 + X3 + X4 + age_s,
+    model_var = "assigned_treatment",
+    glm_function = "glm",
+    use_sample_weights = TRUE,
+    quiet = TRUE)
R> summary(PP_result,  digits = 2)
\end{Sinput}
\begin{Soutput}
Trial Emulation Outcome Model

Outcome model formula:
outcome ~ assigned_treatment + trial_period + I(trial_period^2) + 
    followup_time + I(followup_time^2) + X1 + X2 + X3 + X4 + 
    age_s

Coefficent summary (robust):
              names estimate robust_se  2.5%  97.5%     z p_value
        (Intercept)   -4.477     0.685 -5.82 -3.134 -6.54 6.3e-11
 assigned_treatment   -1.377     0.472 -2.30 -0.452 -2.92 0.00351
       trial_period    0.138     0.247 -0.35  0.621  0.56 0.57742
  I(trial_period^2)   -0.112     0.077 -0.26  0.039 -1.45 0.14577
      followup_time    0.351     0.316 -0.27  0.970  1.11 0.26606
 I(followup_time^2)   -0.059     0.056 -0.17  0.051 -1.06 0.29082
                 X1   -0.440     0.546 -1.51  0.630 -0.81 0.41991
                 X2   -0.082     0.157 -0.39  0.226 -0.52 0.60098
                 X3    1.209     0.392  0.44  1.978  3.08 0.00206
                 X4    0.841     0.234  0.38  1.301  3.59 0.00033
              age_s    0.266     0.510 -0.73  1.266  0.52 0.60142

PP_result$model contains the fitted glm model object.
PP_result$robust$matrix contains the full robust covariance matrix.
\end{Soutput}
\end{Schunk}
The returned object \code{PP_result} includes a list of the summaries of the fitted model, the parameter estimates with robust standard errors and the sandwich variance matrix. From the model summary,  we can see that the estimated coefficient of assigned treatment is $-1.38$ with 95\% confidence interval $[-2.30, -0.45]$, which is statistically significant. This is in contrast to the estimated coefficient of assigned treatment in the intention-to-treat analysis, which is non-significant.

\begin{figure}[t!]
\centering
\includegraphics[width=\textwidth]{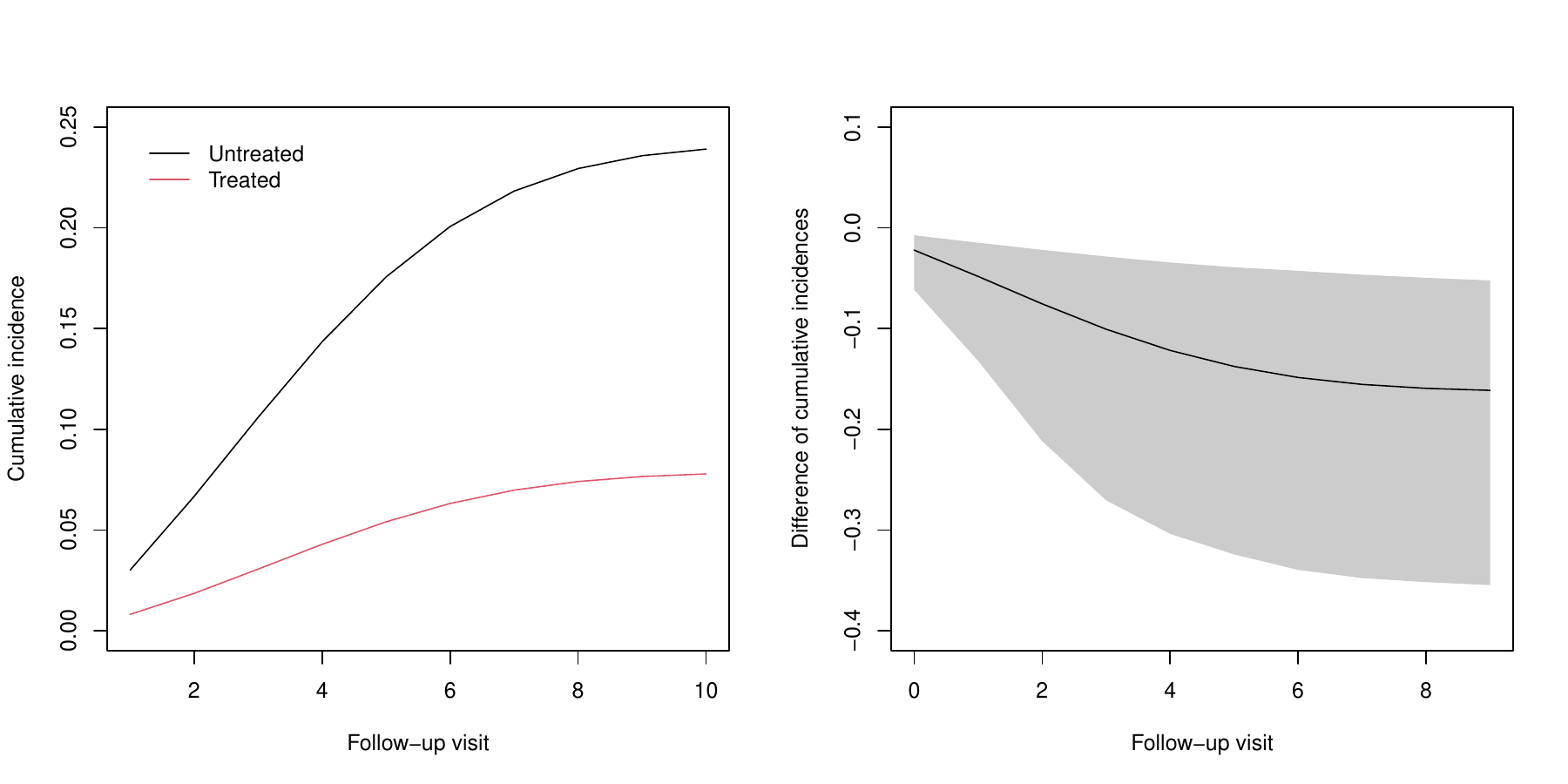}
\caption{\label{fig:PP} Predicted cumulative incidences for untreated and treated and their differences with point-wise 95\% confidence intervals as the marginal per-protocol effect for the trial 0 population. }
\end{figure}

Again we can use the \fct{predict} function to obtain the marginal per-protocol effect in terms of cumulative incidence differences in a target population (in this case, the trial 0 population), following the same steps as in  Section~\ref{ITTillustration}. From Figure~\ref{fig:PP}, it is easy to see that the marginal per-protocol effect in the trial 0 population is also statistically significant, as the point-wise 95\% confidence bands of the cumulative incidence differences do not cover zero.
\begin{Schunk}
\begin{Sinput}
R> new_data <- rbind(data.table::as.data.table(prep_PP_data$data_template),
+    trial_0)
R> set.seed(20222024)
R> PP_cuminc <- predict(PP_result, new_data, predict_times = c(0:9))
R> pdf("PP_cuminc.pdf", width = 12, height = 6)
R> par(mfrow = c(1,2))
R> plot(PP_cuminc$assigned_treatment_0$cum_inc, ty = "l",
+    ylab = "Cumulative incidence", xlab = "Follow-up visit",
+    ylim = c(0, 0.25))
R> lines(PP_cuminc$assigned_treatment_1$cum_inc, col = 2)
R> legend(1, 0.25, legend = c('Untreated', 'Treated'), col = 1:2,
+    lty = c(1,1), bty='n')
R> followup <- PP_cuminc$difference$followup_time
R> cum_inc_diff <- PP_cuminc$difference$cum_inc_diff
R> LL <- PP_cuminc$difference$`2.5%`
R> UL <- PP_cuminc$difference$`97.5%`
R> plot(followup, cum_inc_diff, ty = "l",
+    ylab = "Difference of cumulative incidences",
+    xlab = "Follow-up visit", ylim = c(-0.4, 0.1))
R> polygon(c(followup, rev(followup)), c(LL, rev(UL)),
+    col = 'grey80', border = NA)
R> lines(followup, cum_inc_diff)
\end{Sinput}
\end{Schunk}

\newpage

\section{Summary and discussion} \label{sec:summary}
Target trial emulation is an important framework for conducting causal analyses using observational data.
This paper introduces the \pkg{TrialEmulation} package and demonstrates its functionality with a simulated data example for emulating a sequence of target trials with a time-to-event outcome.
\pkg{TrialEmulation} offers point estimation and inference for marginal intention-to-treat and per-protocol effects in terms of cumulative incidence differences for any target population of interest.
To enable target trial emulation with a large amount of data, \pkg{TrialEmulation} allows data processing by chunks and case-control sampling, in order to create a data set that is manageable in \proglang{R}.

While \pkg{TrialEmulation} focuses on emulating a sequence of target trials, it can certainly emulate a single target trial if users provide data such that the eligibility criteria are satisfied at most once for an individual, and the enrolment visits are coded as the same (e.g., 0) across individuals and the values for follow-up visits/periods saved in the variable \code{period} are coded accordingly using the enrolment visits as the index visits.

The \pkg{TrialEmulation} package is under continuous development to include other useful components for causal analyses in emulated trials. Specifically, we will add diagnostic tools, such as covariate balance summaries for the weight models \citep{Jackson2019}.   Bootstrap confidence intervals will also be included in future versions of \pkg{TrialEmulation}, as it is well known that the sandwich variance estimator with fixed weights can be conservative \citep{Seaman2013,Austin2016,Austin2022}. We are also exploring accommodating competing events in the analyses of emulated trials \citep{Young2020}.


\section*{Computational details}

\pkg{TrialEmulation}~0.0.3.8 package (license: Apache License ($\ge$ 2)) was built on \proglang{R} ($\ge$ 3.5.0)  and dependent on the packages	 \pkg{broom} ($\ge$ 0.7.10),  \pkg{checkmate},  \pkg{data.table} ($\ge$ 1.9.8),  \pkg{formula.tools},  \pkg{mvtnorm},  \pkg{parglm},  \pkg{Rcpp},  \pkg{sandwich}. Package \pkg{splines} is not a dependent package of \pkg{TrialEmulation} but was used to specify the splines terms in our data example. All these packages used are available from the Comprehensive R Archive Network (CRAN) at https://CRAN.R-project.org/.

\section*{Acknowledgments}


The authors would like to thank Juliette Limozon for testing earlier versions of the \pkg{TrialEmulation} package. LS, RR, and SS were funded by the UK Medical Research Council  [Unit programme numbers: MC\_UU\_00002/10, MC\_UU\_00002/15].
For the purpose of open access, the author has applied a Creative Commons Attribution (CC BY) licence to any Author Accepted Manuscript version arising from this submission.

\bibliography{refs_TE}

\begin{thebibliography}{23}
\newcommand{\enquote}[1]{``#1''}
\providecommand{\natexlab}[1]{#1}
\providecommand{\url}[1]{\texttt{#1}}
\providecommand{\urlprefix}{URL }
\expandafter\ifx\csname urlstyle\endcsname\relax
  \providecommand{\doi}[1]{doi:\discretionary{}{}{}#1}\else
  \providecommand{\doi}{doi:\discretionary{}{}{}\begingroup
  \urlstyle{rm}\Url}\fi
\providecommand{\eprint}[2][]{\url{#2}}

\bibitem[{Austin(2016)}]{Austin2016}
Austin PC (2016).
\newblock \enquote{{Variance Estimation when Using Inverse Probability of
  Treatment Weighting (IPTW) with Survival Analysis}.}
\newblock \emph{Statistics in Medicine}, \textbf{35}(30), 5642--5655.
\newblock \doi{10.1002/sim.7084}.

\bibitem[{Austin(2022)}]{Austin2022}
Austin PC (2022).
\newblock \enquote{{Bootstrap vs Asymptotic Variance Estimation when Using
  Propensity Score Weighting with Continuous and Binary Outcomes}.}
\newblock \emph{Statistics in Medicine}.
\newblock \doi{10.1002/sim.9519}.

\bibitem[{Breslow and Day(1987)}]{Breslow1987}
Breslow N, Day N (1987).
\newblock \emph{{Statistical Methods in Cancer Research. Volume II--The Design
  and Analysis of Cohort Studies}}.
\newblock 82. IARC scientific publications.

\bibitem[{Christoffersen \emph{et~al.}(2022)Christoffersen, Williams, and
  {Boost developers}}]{Christoffersen2022}
Christoffersen B, Williams A, {Boost developers} (2022).
\newblock \emph{\pkg{parglm}: Parallel GLM}.
\newblock \proglang{R} package version 0.1.7,
  \urlprefix\url{https://cran.r-project.org/package=parglm}.

\bibitem[{Cole and Hern{\'{a}}n(2008)}]{Cole2008}
Cole SR, Hern{\'{a}}n MA (2008).
\newblock \enquote{{Constructing Inverse Probability Weights for Marginal
  Structural Models}.}
\newblock \emph{American Journal of Epidemiology}, \textbf{168}(6), 656--664.
\newblock \doi{10.1093/aje/kwn164}.

\bibitem[{Danaei \emph{et~al.}(2013)Danaei, Rodr{\'{i}}guez, Cantero, Logan,
  and Hern{\'{a}}n}]{Danaei2013}
Danaei G, Rodr{\'{i}}guez LAG, Cantero OF, Logan R, Hern{\'{a}}n MA (2013).
\newblock \enquote{{Observational Data for Comparative Effectiveness Research:
  An Emulation of Randomised Trials of Statins and PrimaryPrevention of
  Coronary Heart Disease}.}
\newblock \emph{Statistical Methods in Medical Research}, \textbf{22}(1),
  70--96.
\newblock \doi{10.1177/0962280211403603}.

\bibitem[{Gran \emph{et~al.}(2010)Gran, Røysland, Wolbers, Didelez, Sterne,
  Ledergerber, Furrer, von Wyl, and Aalen}]{Gran2010}
Gran JM, Røysland K, Wolbers M, Didelez V, Sterne JAC, Ledergerber B, Furrer
  H, von Wyl V, Aalen OO (2010).
\newblock \enquote{{A Sequential Cox Approach for Estimating the Causal Effect
  of Treatment in the Presence of Time-dependent Confounding Applied to Data
  from the Swiss HIV Cohort Study}.}
\newblock \emph{Statistics in Medicine}, \textbf{29}(26), 2757--2768.
\newblock \doi{10.1002/sim.4048}.

\bibitem[{Hern{\'{a}}n \emph{et~al.}(2008)Hern{\'{a}}n, Alonso, Logan,
  Grodstein, Michels, Willett, Manson, and Robins}]{Hernan2008}
Hern{\'{a}}n MA, Alonso A, Logan R, Grodstein F, Michels KB, Willett WC, Manson
  JE, Robins JM (2008).
\newblock \enquote{{Observational Studies Analyzed like Randomized Experiments:
  An Application to Postmenopausal Hormone Therapy and Coronary Heart
  Disease}.}
\newblock \emph{Epidemiology}, \textbf{19}(6), 766--779.
\newblock \doi{10.1097/EDE.0b013e3181875e61}.

\bibitem[{Hern\'{a}n \emph{et~al.}(2000)Hern\'{a}n, Brumback, and
  Robins}]{Hernan2000}
Hern\'{a}n MA, Brumback B, Robins JM (2000).
\newblock \enquote{{Marginal Structural Models to Estimate the Causal Effect of
  Zidovudine on the Survival of HIV-Positive Men}.}
\newblock \emph{Epidemiology}, \textbf{11}(5), 561--570.

\bibitem[{Hern{\'{a}}n \emph{et~al.}(2009)Hern{\'{a}}n, McAdams, McGrath,
  Lanoy, and Costagliola}]{Hernan2009}
Hern{\'{a}}n MA, McAdams M, McGrath N, Lanoy E, Costagliola D (2009).
\newblock \enquote{{Observation Plans in Longitudinal Studies with Time-varying
  Treatments}.}
\newblock \emph{Statistical Methods in Medical Research}, \textbf{18}(1),
  27--52.
\newblock \doi{10.1177/0962280208092345}.

\bibitem[{Hern{\'{a}}n and Robins(2016)}]{Hernan2016}
Hern{\'{a}}n MA, Robins JM (2016).
\newblock \enquote{{Using Big Data to Emulate a Target Trial When a Randomized
  Trial Is Not Available}.}
\newblock \emph{American Journal of Epidemiology}, \textbf{183}(8), 758--764.
\newblock \doi{10.1093/aje/kwv254}.

\bibitem[{Jackson(2019)}]{Jackson2019}
Jackson JW (2019).
\newblock \enquote{{Diagnosing Covariate Balance Across Levels of
  Right-Censoring Before and After Application of
  Inverse-Probability-of-Censoring Weights}.}
\newblock \emph{American Journal of Epidemiology}, \textbf{188}(12),
  2213--2221.
\newblock ISSN 0002-9262.
\newblock \doi{10.1093/aje/kwz136}.

\bibitem[{Keogh \emph{et~al.}(2023)Keogh, Gran, Seaman, Davies, and
  Vansteelandt}]{Keogh2021}
Keogh RH, Gran JM, Seaman SR, Davies G, Vansteelandt S (2023).
\newblock \enquote{Causal inference in survival analysis using longitudinal
  observational data: Sequential trials and marginal structural models.}
\newblock \emph{Statistics in Medicine}, \textbf{42}(13), 2191--2225.
\newblock \doi{10.1002/sim.9718}.

\bibitem[{Logan \emph{et~al.}(2011)Logan, Danaei, and Hern{\'{a}}n}]{Logan2011}
Logan R, Danaei G, Hern{\'{a}}n MA (2011).
\newblock \emph{{\pkg{Initiators}: \proglang{SAS} Macro to Estimate the
  Observational Analogs of the Intention-to-treat and Per-protocol Effects of
  Hypothetical Treatment Stategies}}.
\newblock \urlprefix\url{https://causalab.sph.harvard.edu/software/}.

\bibitem[{Mandel(2013)}]{Mandel2013}
Mandel M (2013).
\newblock \enquote{{Simulation-based Confidence Intervals for Functions with
  Complicated Derivatives}.}
\newblock \emph{The American Statistician}, \textbf{67}(2), 76--81.
\newblock \doi{10.1080/00031305.2013.783880}.

\bibitem[{Murray \emph{et~al.}(2021)Murray, Caniglia, and Petito}]{Murray2021}
Murray EJ, Caniglia EC, Petito LC (2021).
\newblock \enquote{{Causal Survival Analysis: A guide to Estimating
  Intention-to-treat and Per-protocol Effects from Randomized Clinical Trials
  with Non-adherence}.}
\newblock \emph{Research Methods in Medicine {\&} Health Sciences},
  \textbf{2}(1), 39--49.
\newblock \doi{10.1177/2632084320961043}.

\bibitem[{Rezvani \emph{et~al.}(2024)Rezvani, Gravestock, and Su}]{Rezvani2022}
Rezvani R, Gravestock I, Su L (2024).
\newblock \emph{{\pkg{TrialEmulation}: Causal Analysis of Observational
  Time-To-Event Data}}.
\newblock \proglang{R} package version 0.0.3.8,
  \urlprefix\url{https://cran.r-project.org/package=TrialEmulation}.

\bibitem[{Robins \emph{et~al.}(2000)Robins, Hern\'{a}n, and
  Brumback}]{Robins2000}
Robins JM, Hern\'{a}n MA, Brumback B (2000).
\newblock \enquote{{Marginal Structural Models and Causal Inference in
  Epidemiology}.}
\newblock \emph{Epidemiology}, \textbf{11}(5), 550--560.
\newblock \doi{10.1097/00001648-200009000-00011}.

\bibitem[{{\proglang{SAS} Institute Inc.}(2008)}]{SAS-STAT}
{\proglang{SAS} Institute Inc} (2008).
\newblock \emph{\proglang{SAS/STAT} Software, Version~9.2}.
\newblock Cary, NC.
\newblock \urlprefix\url{https://www.sas.com/}.

\bibitem[{Seaman and White(2013)}]{Seaman2013}
Seaman SR, White IR (2013).
\newblock \enquote{{Review of Inverse Probability Weighting for Dealing with
  Missing Data}.}
\newblock \emph{Statistical Methods in Medical Research}, \textbf{22}(3),
  278--295.
\newblock \doi{10.1177/0962280210395740}.

\bibitem[{Young \emph{et~al.}(2020)Young, Stensrud, {Tchetgen Tchetgen}, and
  Hern{\'{a}}n}]{Young2020}
Young JG, Stensrud MJ, {Tchetgen Tchetgen} EJ, Hern{\'{a}}n MA (2020).
\newblock \enquote{{A Causal Framework for Classical Statistical Estimands in
  Failure-time Settings with Competing Events}.}
\newblock \emph{Statistics in Medicine}, \textbf{39}(8), 1199--1236.
\newblock \doi{10.1002/sim.8471}.

\bibitem[{Young and Tchetgen~Tchetgen(2014)}]{Young2014}
Young JG, Tchetgen~Tchetgen EJ (2014).
\newblock \enquote{{Simulation from a Known Cox MSM using Standard Parametric
  Models for the G-formula}.}
\newblock \emph{Statistics in Medicine}, \textbf{33}(6), 1001--1014.
\newblock \doi{10.1002/sim.5994}.

\bibitem[{Zeileis \emph{et~al.}(2020)Zeileis, K\"ollm, and
  Graham}]{Zeileis2020}
Zeileis A, K\"ollm S, Graham N (2020).
\newblock \enquote{Various Versatile Variances: An Object-Oriented
  Implementation of Clustered Covariances in {R}.}
\newblock \emph{Journal of Statistical Software}, \textbf{95}(1), 1--36.
\newblock \doi{10.18637/jss.v095.i01}.

\end{thebibliography}


\begin{appendix}

\section{Details of causal estimands and assumptions}\label{assumptions}

Following the notation in Section~\ref{singletrial}, we describe the causal estimands and their identifying assumptions in a single emulated trial setting.
Recall that we use overbars to denote the history of a random variable,
for example, $\overline{Y}_k = (Y_0, \ldots, Y_k) $ is the history of the outcome up to immediately before time  $t_{k+1}$. 
We shall denote the time immediately before time $t_{k+1}$ as $t_{k+1}^-$.
In addition, we denote the future of a random variable up to the follow-up time of interest using underbars, for example, $\underline{Y}_{k+1} =(Y_{k+1}, \ldots,  Y_K)$.

\subsection{The intention-to-treat effect}\label{ITTassumption}
The intention-to-treat effect is the effect of treatment randomisation on the outcome of interest, which is the primary effect of interest in RCTs.
As we elaborate on the scenario with dependent censoring, we introduce 
 the potential outcome variables $Y_{k}^{a,\overline{c}=0}$  ($a=0,1$)  as follows. For each individual, $Y_{k}^{a,\overline{c}=0}$ is the indicator of whether the event of interest would have  occurred during the time interval $[0, t_{k+1})$ if the individual, possibly contrary to fact, had been assigned to the treatment $a$ and if there had been a hypothetical intervention that had eliminated all censoring during the follow-up period up to  $t_{k+1}^-$ (denoted as $\overline{c}=0$) \citep{Young2020}.
Similarly, we define the potential time-varying covariates $L_{k}^{a,\overline{c}=0}$ if the individual had been assigned to the treatment $a$ and if there had been a hypothetical intervention that had eliminated all censoring during the follow-up period up to  $t_{k+1}^-$.

The marginal intention-to-treat effect can be defined in terms of differences between cumulative incidences.
$$
  \mbox{Pr}(Y_{k}^{a=1,\overline{c}=0}=1)-\mbox{Pr}(Y_{k}^{a=0,\overline{c}=0}=1),
$$ or equivalently, the differences between  survival probabilities,
$$
  \mbox{Pr}(Y_{k}^{a=0,\overline{c}=0}=0)-\mbox{Pr}(Y_{k}^{a=1,\overline{c}=0}=0).
$$

To identify the marginal  intention-to-treat  effect, \pkg{TrialEmulation} makes the following assumptions:

\begin{enumerate}

\item[(1)] \emph{No interference.} It is assumed that the potential outcome variables for a given individual, $Y_{k}^{a,\overline{c}=0}$,
do not depend on the treatments assigned to any other individuals. 

\item[(2)] \emph{Consistency}: If $A_0=a$ and $\overline{C}_{k}=0$, then $\overline{L}_{k}=\overline{L}_{k}^{a,\overline{c}=0}$ and $\overline{Y}_{k}=\overline{Y}_{k}^{a,\overline{c}=0}$.  
This assumption requires that, if an individual has data consistent with the intervention indexing the potential outcome and covariates up to  $t_{k+1}^-$, then his/her observed outcomes and covariates up to $t_{k+1}^-$ equal his/her potential outcomes and covariates under that intervention.
A check for the consistency assumption is to ensure that the intervention is well-defined, that is, each individual is receiving the same version of the treatment.

\item[(3)] \emph{Positivity of treatment assignment.} If $f(v, l_0)>0$,  then $\mbox{Pr}(A_0 =a \mid V, L_0)>0$ for $a=0,1$, where $f(v, l_0)$ is the joint density of $V$ and $L_0$. This assumption requires that each individual has a non-zero probability of being assigned to treatment/non-treatment within any possibly observed levels of  $V$ and $L_0$.

\item[(4)] \emph{No unmeasured confounding of  treatment assignment. }   Given $V$ and $L_0$, the potential outcomes $\overline{Y}_K^{a, \overline{c}=0}$ are independent of the treatment assignment at the baseline, i.e., $\overline{Y}_K^{a, \overline{c}=0} \indep A_0 \mid V,  L_0$.  This means that there is no unmeasured confounding for treatment assignment and the outcome event, conditional on   $V$ and $L_0$.

 \item[(5)] \emph{Positivity of being uncensored}:  For $k=1, \ldots,K$, if $f_{C_{k-1}, Y_{k}, \overline{A}_{k}, V, \overline{L}_{k}, }(0, 0, \overline{a}_{k}, v,  \overline{l}_{k}) > 0$, $\mbox{Pr} (C_{k}=0 \mid {C}_{k-1}={Y}_{k}=0, \overline{A}_{k} = \overline{a}_{k}, V=v,  \overline{L}_{k}=\overline{l}_{k})>0$, where \\$f_{C_{k-1}, Y_{k}, \overline{A}_{k}, V, \overline{L}_{k}, }(0, 0, \overline{a}_{k}, v,  \overline{l}_{k})$ is the joint density of $(C_{k-1}, Y_{k}, \overline{A}_{k}, V, \overline{L}_{k} )$ evaluated at $(0, 0, \overline{a}_{k}, v,  \overline{l}_{k})$.  This means that, given the measured past, each individual has non-zero probability of remaining uncensored by $t_{k+1}^-$.

\item[(6)] \emph{Sequential ignorability of censoring}:  For $k=1, \ldots,K$,  $\underline{Y}_{k+1}^{a,\overline{c}=0} \indep C_{k} \mid \overline{C}_{k-1}=\overline{Y}_{k}=0,  \overline{A}_{k} = \overline{a}_{k}, V= v, \overline{L}_{k}=\overline{l}_{k}$, where $v$ and $\overline{l}_{k}$ are  some realisations of  $V$ and $\overline{L}_{k}$, respectively. This assumption means that, at each follow-up visit, given the treatment history, baseline covariates and covariate history, censoring is independent of future potential outcomes had everyone been assigned $a$ and no censoring occurred.

\end{enumerate}

\subsection{The per-protocol effect}
The per-protocol effect is the effect of adherence to the assigned treatment strategy defined by the study protocol, in other words, the effect of sustained treatments.
For each individual, $Y_{k}^{\overline{a}_k,\overline{c}=0}$ is the indicator of whether the event of interest has  occurred by $t_{k+1}^-$,  if the individual, possibly contrary to fact, had taken treatment $\overline{a}_k$ up to time $t_k$ and no censoring had occurred ($\overline{c}=0$).
We consider two values of $\overline{a}_k$, viz.\ $\overline{a}_k= (1, \ldots, 1)$ or $\overline{a}_k= (0, \ldots, 0)$.
These correspond to starting treatment/non-treatment at baseline and adhering to this treatment/non-treatment until at least time $t_k$.\\

The marginal per-protocol effect can be defined in terms of differences between  cumulative incidences,
$$
  \mbox{Pr}(Y_{k}^{\overline{a}_k=(1, \ldots, 1),\overline{c}=0}=1)-\mbox{Pr}(Y_{k}^{\overline{a}_k=(0, \ldots, 0),\overline{c}=0}=1),
$$ or differences between  survival probabilities,
$$
  \mbox{Pr}(Y_{k}^{\overline{a}_k=(0, \ldots, 0),\overline{c}=0}=0)-\mbox{Pr}(Y_{k}^{\overline{a}_k=(1, \ldots, 1),\overline{c}=0}=0).
$$

To identify the per-protocol effects, \pkg{TrialEmulation} makes the following assumptions.
Note that in these assumptions, we are only considering two values of $\overline{a}_k$: $\overline{a}_k= (1, \ldots, 1)$ and $\overline{a}_k= (0, \ldots, 0)$.
  \begin{enumerate}

\item[(1)] \emph{No interference.} It is assumed that the potential outcome variables for a given individual, $Y_{k}^{\overline{a}_k,\overline{c}=0}$,
do not depend on the treatment assignment and adherence of any other individuals.

\item[(2)] \emph{Consistency}: If $\overline{A}_k=\overline{a}_k$ and $\overline{C}_{k}=0$, then $\overline{L}_{k}=\overline{L}_{k}^{\overline{a}_k,\overline{c}=0}$ and $\overline{Y}_{k}=\overline{Y}_{k}^{\overline{a}_k,\overline{c}=0}$.
This means that if an individual has treatment history up to time $t_{k+1}^-$ consistent with the intervention indexing the potential outcome and is not censored, then his/her observed outcomes and covariates equal his/her potential outcomes and covariates under that intervention.

 \item[(3)] \emph{Positivity of treatment assignment and  adherence.} For $k=0, \ldots,K$, if \\$f_{Y_{k-1}, C_{k-1},\overline{A}_{k-1}, V, \overline{L}_{k}, }(0, 0, \overline{a}_{k-1}, v, \overline{l}_{k}) > 0$, then $\mbox{Pr}(A_k=a_{k-1} \mid  \overline{Y}_{k-1}=\overline{C}_{k-1}=0,  \overline{A}_{k-1}=\overline{a}_{k-1}, V, \overline{L}_{k}=\overline{l}_{k})>0$, 
 where $f_{Y_{k-1}, C_{k-1},\overline{A}_{k-1}, V, \overline{L}_{k}, }(0, 0, \overline{a}_{k-1}, v, \overline{l}_{k}) > 0$ is the joint density of $(Y_{k-1}, C_{k-1}, \overline{A}_{k-1}, V, \overline{L}_{k})$ evaluated at $(0,0, \overline{a}_{k-1}, v, \overline{l}_{k})$.  By convention, we have $\overline{a}_{-1}=\emptyset$, $\overline{Y}_{-1}=\overline{C}_{-1}=0$.  This implies, in particular, that individuals in each subgroup defined by baseline covariates have non-zero probability of being assigned to the treatment/non-treatment at $t_0$. Moreover,  individuals in each subgroup defined
 by the  treatment adherence  ($\overline{a}_{k-1}= (1, \ldots, 1)$ or $\overline{a}_{k-1}= (0, \ldots, 0)$) and covariate history have a non-zero probability of being adherent
at $t_k$. 

\item[(4)] \emph{Conditional exchangeability of treatment assignment and adherence}: For $k=0,\ldots, K$, ${Y}_k^{\overline{a}_k, \overline{c}=0} \indep A_k \mid \overline{A}_{k-1}=\overline{a}_{k-1}, V=v,  \overline{L}_k=\overline{l}_k, \overline{Y}_{k-1}=0$, for $\overline{a}_{k-1}= (1, \ldots, 1)$ or $\overline{a}_{k-1}= (0, \ldots, 0)$, and for all $v$ and $\overline{l}_{k}$.
This assumption means that, conditional on baseline covariates,  the treatment and covariate histories, treatment assignment and adherence are independent of the potential outcomes. That is, there is no unmeasured confounding between treatment assignment,  time-varying treatment adherence and event of interest.

\item[(5)] \emph{Positivity of being uncensored}: For $k=1, \ldots,K$, if $f_{C_{k-1}, Y_{k}, \overline{A}_{k}, V, \overline{L}_{k}, }(0, 0, \overline{a}_{k}, v,  \overline{l}_{k}) > 0$, $\mbox{Pr} (C_{k}=0 \mid {C}_{k-1}={Y}_{k}=0, \overline{A}_{k} = \overline{a}_{k}, V=v,  \overline{L}_{k}=\overline{l}_{k})>0$, where \\$f_{C_{k-1}, Y_{k}, \overline{A}_{k}, V, \overline{L}_{k}, }(0, 0, \overline{a}_{k}, v,  \overline{l}_{k})$ is the joint density of $(C_{k-1},Y_{k}, \overline{A}_{k}, V, \overline{L}_{k} )$ evaluated at $(0, 0, \overline{a}_{k}, v,  \overline{l}_{k})$. Again this assumption requires that,  given the measured past, each individual has a non-zero probability of remaining uncensored by $t_{k+1}^-$.

\item[(6)] \emph{Sequential ignorability of censoring}:  For $k=1, \ldots,K$, $\underline{Y}_{k+1}^{\overline{a}_{k},\overline{c}=0} \indep C_{k} \mid \overline{C}_{k-1}=\overline{Y}_{k}=0,  \overline{A}_{k}=\overline{a}_{k}, V=v, \overline{L}_{k}=\overline{l}_{k} $ for $\overline{a}_{k-1}= (1, \ldots, 1)$ or $\overline{a}_{k-1}= (0, \ldots, 0)$, and for all $v$ and $\overline{l}_{k}$ that have non-zero probability of occurring.
This assumption requires that, at each follow-up visit, given the treatment being adherent to, 
baseline covariates and covariate history, censoring is independent of future potential outcomes had everyone been adhering to the assigned treatments and no censoring occurs.

 \end{enumerate}

\section{Details of optional arguments in functions}\label{moreoptions}

Tables~\ref{tab:prep}-\ref{tab:sampling} describe other optional arguments in the functions \fct{data\_preparation}, \\\fct{trial\_msm},  \fct{case\_control\_sampling\_trials} and \fct{predict}.
\begin{table}[hbp]
    \centering
     \caption{Optional arguments in the  \fct{data\_preparation} function}
    \label{tab:prep}
    \begin{tabular}{p{4.3cm}p{9cm}}
    \hline
       \hline
    Argument  & Description \\
      \hline
 \code{first\_period}, \code{last\_period} & Set the first time period and the last time period as trial baselines to start expanding the data.     \\
&\\


&\\
  \code{eligible\_wts_0}, \code{eligible\_wts_1}& Exclude some observations from the  models for the inverse probability of treatment weights. For example, if it is assumed that an individual will stay on treatment for at least 2 visits, the first 2 visits after treatment initiation by definition have a probability of staying on the treatment of 1.0 and should thus be excluded from the weight models for those who are on treatment at the immediately previous visit. Users can define a variable that indicates that these 2 observations are ineligible for the weight model for those who are on treatment at the immediately previous visit and add the variable name in the argument   \code{eligible\_wts_1}. Similar definitions are applied to \code{eligible\_wts_0} for excluding observations when fitting the models for the inverse probability of treatment weights for those who are \emph{not} on treatment at the immediately previous visit.\\
\code{where\_var} & Specify the variable names that will be used to define subgroup conditions when fitting a marginal structural model for a subgroup of individuals. Need to specify jointly with the argument \code{where\_case} in the  \fct{trial\_msm} function.  \\
\code{quiet = TRUE/FALSE} &  Suppress the printing of progress messages and summaries of the fitted models for calculating the weights.\\
  \hline   \end{tabular}
\end{table}

\begin{table}[htp]
    \centering
\caption{Optional arguments in the  \fct{trial\_msm} function}
    \label{tab:model}

    \begin{tabular}{p{4.8cm}p{9cm}}
    \hline
    \hline
      Argument & Description \\
      \hline
 \code{first_followup}, \code{last_followup} & Set part of follow-up visits to be included in the marginal structural model for the outcome event. \code{first\_followup}, \code{last\_followup} are desired first and last  follow-up visits (inclusive). \\
              &   \\
              \code{analysis_weights = c("asis", "unweighted", "p99", "weight_limits")} & Choose which type of weights to be used for fitting the marginal structural model for the outcome event.  \code{"asis"}: use the weights as calculated in \fct{data\_preparation};
              \code{"p99"}: use weights truncated at the 1st and 99th percentiles (based on the distribution of weights in the entire sample);
              \code{"weight_limits"}: use weights truncated at the values specified in the argument \code{weight_limits};
              \code{"unweighted"}:   no weighting is applied. That is, set weights to 1. \\
        &   \\
\code{weight_limits} &
Lower and upper limits to truncate weights, given as c(lower, upper).\\

             &   \\
 \code{where_case} &  Define conditions using variables specified in \code{where\_var} when fitting a marginal structural model for a subgroup of the individuals. For example, if \code{where\_var= "age"}, \code{where_case = "age >= 30"} will only fit the pooled logistic model to the subset of inividivuals who are 30 years old or above. \\
\code{quiet = TRUE/FALSE} &  Suppress the printing of progress messages and the summary of the fitted pooled logistic model with and without robust standard errors.\\
\hline
\end{tabular}
\end{table}

\begin{table}[htp]
    \centering
     \caption{Optional arguments in  \fct{case\_control\_sampling\_trials} and \fct{predict} }
    \label{tab:sampling}
    \begin{tabular}{p{4.8cm}p{9cm}}
    \hline
       \hline
 Argument & Description \\
\hline
\textbf{\fct{case\_control\_sampling\_trials} }& \\
~~~~~~\code{subset_condition} &  Expression to subset the expanded data before applying case-control sampling) \\
~~~~~~\code{sort = TRUE/FALSE} & Sort data before applying case-control sampling to make sure that the resulting data are identical when sampling from the expanded data created with \code{separate_files = TRUE} or \code{separate_files = FALSE}.\\
&\\
\fct{predict}& \\
~~~~~~\code{samples = 100} & Number of samples used to construct the simulation-based confidence intervals.\\
~~~~~~\code{type = c("cum_inc",} & Type of measures to be predicted.  \\
~~~~~~~~~~~~~~\code{"survival")} & \code{type = "cum_inc"}: predict cumulative incidences; \code{type = "survival"}: predict survival probabilities.\\
\hline
\end{tabular}
\end{table}

\newpage
\section{Algorithm to simulate data in Section 4} \label{app:simdata}

\begin{figure}[htp]
		\centering
  \caption{\label{fig_sim} The relationship between time-varying variables in the simulation set-up.}
		\begin{tikzpicture}
		\tikzstyle{connect}=[->,shorten >=1pt,node distance=10cm,semithick]
		\node (1)  {$A_{j-1}$};
		\node (2) [right=1cm of 1] {$X_{1, j}, X_{2, j}$};
		\node (3) [right=1cm of 2] {$A_j$};
		\node (4) [below =1cm of 2] {$C_{j}$};
		\node (7) [below =1cm of 3] {$Y_{j}$};

		\path (1) edge [connect] (2);
		\path (1) edge [connect, bend left] (3);
		\path (2) edge [connect] (3);
		\path (1) edge [connect, bend right] (4);
        \path (2) edge [connect] (4);
		\path (2) edge [connect] (7);
		\path (3) edge [connect] (7);

		\end{tikzpicture}

	\end{figure}
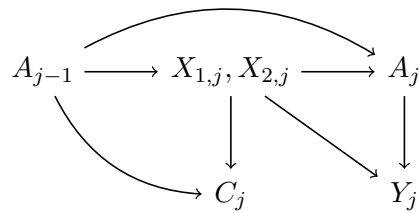

We follow the algorithm described in \cite{Young2014} to simulate data with time-varying confounding. Details of the data generating mechanism are presented in Table~\ref{DGM}. Figure~\ref{fig_sim} also describes the relationships between the time-varying variables in the simulated data, where time-invariant variables $X_3$ and $X_4$ are omitted for clearer exposition.

\begin{table}[htp]
\centering
\caption{\label{DGM} Summary of data generating mechanism of the simulated data}
\small
\begin{tabular}{l l}
 \hline
 \hline
 Number of individuals:&  $n=1000$   \\
Baseline and follow-up visits: & $j = 0,\ldots, 9$  \\
 & \\
 Time-varying covariates &  $X_{1,0} \sim \text{Bernoulli}(0.5)$ \\
                         &  $X_{1,j} \sim \text{Bernoulli}(p_j)$, $\rmn{logit} (p_j) = - A_{j-1}$  \\
 & $ X_{2,0} \sim N(0,1)$ \\
 & $ X_{2,j} \sim N(- 0.3 A_{j-1},1)$ \\
 & $\text{age}_0 \sim N(35,12^2)$\\
 &  $\text{age}_j = \text{age}_{j-1} + 1$ \\
 &  standarized age: $\text{age}_j^s = (\text{age}_j - 35)/12$ \\
 & \\
 Time-invariant covariates & $X_3 \sim \text{Bernoulli}(0.5)$ \\
 &  $X_4 \sim N(0,1)$ \\
 & \\
 Treatment & $A_{j} \sim \text{Bernoulli}(\pi_j)$\\
 &$\rmn{logit} (\pi_j) = A_{j-1} + 0.5 X_{1,j} + 0.5 X_{2,j} - 0.2 X_3 + X_4 -0.3 \text{age}_j^s$ \\
& \\
 Outcome & $Y_{j} \mid Y_{j-1}=0  \sim \text{Bernoulli}(\lambda_j)$\\
 & $\text{logit}( \lambda_j)  = -5 - 1.2 A_j + 0.5 X_{1,j} + 0.5 X_{2,j} + X_3 + X_4+ 0.5 \text{age}_j^s$ \\
 & \\
 Censoring &   $C_{j} \mid C_{j-1}=0  \sim \text{Bernoulli}(q_j)$\\
 & $\text{logit} (q_j)  =  - 1 - A_{j-1}- 0.5 X_{1,k} +0.5 X_{2,k}- 0.2 X_3 + 0.2X_4 - \text{age}_j^s$ \\
 & \\
 Eligibility & $E_j = 1$ if individuals are 18 years old or older at visit $j$, have not \\
 & received treatment before visit $j$ \\
 & and have not experienced the outcome event before visit $j$; \\
 & $E_j = 0$ otherwise.  \\
  & \\
\multicolumn{2}{l}{Note: Observations before the individuals first become eligible and observations after the outcome } \\
\multicolumn{2}{l}{event or the censoring has occurred are removed from the final simulated data.}\\
 \hline
\end{tabular}

\end{table}

\newpage
 \section{Fitting standard marginal structural models}\label{fitmsm}
Standard marginal structural models for all possible treatment sequences in emulated trials can be fitted by setting \code{estimand_type = "As-Treated"}. Specification of other options is almost identical to per-protocol analyses, except that there is more flexibility to specify the functions of possible treatment sequences.  
If we set \code{model_var = "dose"}, then the data object returned by  \fct{data\_preparation} will include a new variable \code{dose}, which is the cumulative exposure of treatment up to the current follow-up visit. Using the \fct{trial\_msm} function,  a marginal structural model with the linear and quadratic terms of the \code{dose} variable can be fitted,  conditional on the baseline covariates specified in the argument \code{outcome_cov},  functional forms of the follow-up visits specified in the argument \code{include_followup_time} and functional forms of the trial index specified in the argument \code{include_trial_period}.
Currently, the \fct{trial\_msm} function can only include \code{assigned_treatment},  \code{dose}, \code{treatment} and their functions in marginal structural models. 
Other summaries of possible treatment sequences can be created for the data object returned by \fct{data\_preparation}, which can then be used to fit a bespoke marginal structural model using the \fct{glm} function. The \fct{predict} function is not supported when  \code{estimand_type = "As-Treated"} at the moment as pre-specified treatment regimes in terms of the treatment variables specified in  \code{outcome_cov}  are required for predicting marginal treatment effects in a target trial population. 

Fitting standard marginal structural models to observational cohort data using \pkg{TrialEmulation} would require that individuals have the variable \code{eligible} $=1$ only at the baseline visit of the entire  cohort. Since the specification of function arguments and options is almost identical as in emulated trials, we can set \code{estimand_type = "As-Treated"}. Note that we need to set
\code{include_trial_period = ~ 1} because individuals are only eligible at the cohort baseline and no data expansion is performed. 
\end{appendix}


\end{document}